\documentclass[preprint,longtitle,3p,11pt]{elsarticle}

\usepackage{amssymb}

\usepackage{lineno}

\usepackage{epsfig}
\usepackage{epstopdf}
\usepackage{hyperref}
\usepackage{subfigure}
\usepackage{color}
\usepackage{multirow}

\newcommand{\pp}{pp~}
\newcommand{\mpp}{\mathrm{pp}}
\newcommand{\mAA}{\mathrm{AA}}
\newcommand{\sqrts}{\sqrt{s}}
\newcommand{\sqrtsNN}{\sqrt{s_{\rm NN}}}

\newcommand{\gevc}{\mathrm{GeV}/c}

\newcommand{\PbPb}{\mbox{Pb--Pb}}
\newcommand{\pt}{p_{\rm t}}

\newcommand{\Jpsi}{J/$\psi$~}
\newcommand{\dsigmady}{{\rm d} \sigma_{\mpp}/{\rm d} y}

\begin{document}

\begin{frontmatter}



\title{Phenomenological interpolation of the inclusive \Jpsi cross section to proton-proton collisions at $\sqrts=2.76$~TeV and 5.5 TeV}

\author[Torino]{F.~Boss\`u}
\author[add2,add3]{Z.~Conesa~del~Valle}
\author[Cagliari]{A.~de~Falco}
\author[Torino]{M.~Gagliardi}
\author[Doubna,YerPhi]{S.~Grigoryan}
\author[Nantes]{G.~Mart\'{\i}nez~Garc\'{\i}a}

\address[Torino]{Universit\`a degli Studi di Torino and Sezione INFN di Torino, Torino, Italy}
\address[add2]{European Organization for Nuclear Research (CERN), Geneva, Switzerland}
\address[add3]{Institut Pluridisciplinaire Hubert Curien (IPHC), CNRS-IN2P3 UdS, Strasbourg, France}
\address[Cagliari]{Universit\`a di Cagliari and Sezione INFN di Cagliari, Cagliari, Italy}
\address[Doubna]{JINR, Dubna, Russia}
\address[YerPhi]{YerPhI, Yerevan, Armenia}
\address[Nantes]{SUBATECH, Ecole des Mines de Nantes, Universit\'e de Nantes, CNRS-IN2P3, Nantes, France}

\begin{abstract}
We present a study of the inclusive \Jpsi cross section at $\sqrts=2.76$~TeV and $5.5$~TeV. 
The $\sqrts$ dependence of the cross section, rapidity and transverse momentum distributions are evaluated phenomenologically. 
Their knowledge is crucial as a reference for the interpretation of A--A and p--A \Jpsi results at the LHC. 
Our approach is the following: first, we estimate the $\sqrts$ evolution of the $\pt$-integrated \Jpsi  cross section at mid-rapidity; then, we evaluate the rapidity dependence; finally, we study the transverse momentum distribution trend. 
Whenever possible, both theory driven (based on pQCD predictions) and functional form (data driven fits) calculations are discussed. 
Our predictions are compared with the recently obtained results by the ALICE collaboration in \pp collisions at 2.76 TeV.

\end{abstract}
%
%
%

\end{frontmatter}



%

\cleardoublepage
\section{Introduction}
\label{Intro}
High energy heavy-ion collisions are used to study strongly interacting matter under extreme conditions. At sufficiently high collision energies, it is believed that hot and dense deconfined matter, commonly referred to as the Quark-Gluon Plasma (QGP), is formed. With the advent of a new generation of experiments at the CERN Large Hadron Collider (LHC) \cite{LHC} a new energy domain is accessible to study the properties of such a state of matter. 
Many observables have been proposed to probe deconfined matter formed in heavy-ion collisions. 
The production of heavy-quarks and quarkonium at LHC energies is promising since their production rate allows now careful studies of their characteristics ($\pt$ and $y$ distributions, polarization, flow, azimuthal correlations)~\cite{PPR, Armesto08}. 
In particular, the production of charm is expected to be large enough to make regeneration mechanisms possible. The regeneration of \Jpsi mesons due to the recombination of $c$ and $\bar{c}$ quarks in later stage of the heavy-ion collision would be a direct probe of the deconfined QCD matter~\cite{Recombination}. 
One of the experimental methods to quantify the nuclear medium effects in the production of a given observable ($Ob$) is the measurement of the nuclear modification factor ($R^{Ob}_{\mAA}$) in nucleus-nucleus (A--A) collisions, defined as:
\begin{equation}
R^{Ob}_{\mAA}  = \frac{ Y^{Ob}_{\mAA} }{\langle N_{coll} \rangle \, \, Y^{Ob}_{\mpp}}
\end{equation}
where $\langle N_{coll} \rangle$ is the average number of binary nucleon-nucleon collisions\footnote{ 
The average number of binary nucleon-nucleon collisions can be estimated by the product of the average nuclear overlap function (of the nucleus-nucleus collision) and the inelastic proton-proton cross section~\cite{Enterria03}. 
} and  $Y^{Ob}_{\mAA}$ ($Y^{Ob}_{\mpp}$) is the invariant yield of the observable $Ob$ in A--A (pp) collisions at a given (same) center-of-mass energy. 
In the absence of nuclear matter effects, the nuclear modification factor should be equal to unity for experimental observables commonly called \emph{hard probes} (large $\pt$ particles, jets, heavy-flavour, etc). 
A similar factor $R^{Ob}_{\rm pA}$, measured in p--A collisions, is crucial in order to disentangle hot and cold nuclear matter effects in A--A collisions. 

First results on quarkonium production in proton-proton collisions at $\sqrts=7$~TeV have been published by the LHC experiments~\cite{ALICEjpsi10,CMSjpsi10,LHCbjpsi10,ATLASjpsi10} and a successful heavy-ion campaigns took place at the LHC in November 2010 and 2011, when Pb--Pb collisions at the energy per nucleon pair of $\sqrtsNN=2.76$~TeV where delivered to the LHC experiments, for an integrated luminosity of about 9~$\mu$b$^{-1}$.  The nominal LHC energies for \pp and Pb--Pb collisions are $\sqrts=14$~TeV and $\sqrtsNN=5.5$~TeV, respectively.

In this letter, we present an interpolation of the $\pt$-integrated and differential inclusive \Jpsi cross section in \pp collisions to $\sqrts=2.76$~TeV and $5.5$~TeV. We establish a procedure to determine the \Jpsi cross section at arbitrary energy values, which makes possible the calculation of the \Jpsi nuclear modification factors for any colliding system and energy at the LHC\footnote{
As an example, the center-of-mass energy per nucleon pair in p--Pb collisions is 3.5--4~TeV at the present LHC energy, 7~TeV for pp collisions, and of 8.8~TeV at nominal energy. In the present LHC schedule, a p--Pb run is foreseen by the end of 2012.  Operation at nominal energy will be commissioned after 2013.
}. 
The interpolation is done in three steps. The first step is an energy interpolation of the existing $\pt$-integrated \Jpsi production cross section measurements at mid-rapidity (Sec.~\ref{sec:energyinterpolation}). The second step is the evaluation of the energy evolution of the rapidity distribution (Sec.~\ref{sec:rapidityinterpolation}). The calculations are made for the rapidity regions where \Jpsi production is measured at the LHC. The third step is the description of the energy evolution of the transverse momentum distribution (Sec.~\ref{sec:ptinterpolation}). 
Our prediction of the \Jpsi cross section and kinematic distributions at $\sqrts=2.76$~TeV are compared with the results recently obtained by the ALICE collaboration at mid- and forward-rapidity~\cite{Arnaldi11}.

%

\section{Available experimental data}
\label{data}

The \Jpsi interpolation to $\sqrts=2.76$~TeV and $5.5$~TeV has been performed by considering: PHENIX measurements in proton-proton collisions at $\sqrts=200$~GeV in $|y|<2.2$~\cite{PHENIX07}, CDF results in proton-antiproton collisions at $\sqrts=1.96$~TeV in $|y|<0.6$~\cite{CDF05}, and ALICE, CMS and LHCb results in proton-proton collisions at $\sqrts=7$~TeV in various rapidity regions covering $|y|<4.5$~\cite{ALICEjpsi10,CMSjpsi10,LHCbjpsi10}. 
Table~\ref{table1} presents a summary of the exploited data. 
\begin{table}[h]
\begin{center}
\begin{tabular}{||c|cccc||} \hline \hline
Experiment & $\sqrts$  & $y$ range & $\pt$ range    & $\langle \pt \rangle$ \\ 
 ~      &  [GeV] &  ~                 & [GeV]        &  [GeV]  \\ \hline
 PHENIX~\cite{PHENIX07} & 200 & $|y|<0.35$    & $0<\pt<9$ &$1.78\pm 0.14$ \\
 PHENIX~\cite{PHENIX07} & 200 & $1.2<|y|<2.2$ &  $0<\pt<7$ & $1.62\pm 0.10$ \\
 CDF~\cite{CDF05}              & 1960& $|y|<0.6$ & $0<\pt<20$  & $2.51\pm0.09$ \\
 ALICE~\cite{ALICEjpsi10} & 7000 & $-4.0<y<-2.5$ & $0<\pt<8$ & $2.44\pm0.18$ \\ 
 ALICE~\cite{ALICEjpsi10} & 7000 & $|y|<0.9$ & $0<\pt<7$ & $2.73\pm0.44$ \\ 
 CMS~\cite{CMSjpsi10}       & 7000 & $1.6<|y|<2.4$ & $0<\pt<16$ & $2.62\pm0.14$ \\
 LHCb~\cite{LHCbjpsi10}    & 7000 & $2.0<y<4.5$ & $0<\pt<11$ & $2.49\pm0.05$ \\  \hline \hline
\end{tabular}
\end{center}
\caption{Summary of the measurements considered in this work~\cite{ALICEjpsi10,CMSjpsi10,LHCbjpsi10,PHENIX07,CDF05}.}
\label{table1}
\end{table}
In the following, the statistical and systematic uncertainties of the experimental data have been summed in quadrature. 
We assume \Jpsi are produced un-polarized. Indeed, if we suppose polarization does not evolve (or does slowly) with the interaction energy, its influence on the energy interpolation (cross section ratios at different energies or rapidities) should be small for a given experiment. 
Therefore, we neglect the uncertainties related to the polarization.

\subsection{PHENIX results}
\label{sec:dataphenix}

\Jpsi production in \pp collisions at $\sqrts = 200$~GeV has been measured by the PHENIX experiment at the Relativistic Heavy Ion Collider (RHIC) over a rapidity of $-2.2<y<2.2$ and a transverse momentum of $\pt~<~9$~$\gevc$~\cite{PHENIX07}. 
The inclusive differential cross section as a function of rapidity was measured for eleven rapidity bins. 
At mid-rapidity, we have considered the average of the measurements at $|y|<0.35$, that is: 
\newline
$ \textrm{BR}_{ll} \times \dsigmady \big|_{|y|<0.35} = 44.30 \pm 1.40~{\rm (stat.)} \pm 5.10~{\rm (syst.)} \pm 4.50~{\rm (norm.)}$~nb~\cite{PHENIX07}. 
\newline
The mean transverse momentum $\langle \pt \rangle$ has been obtained directly from the experimental data and is reported in Table.~\ref{table1}.

\subsection{CDF results}
\label{sec:datacdf}

The CDF experiment at the Tevatron collider has measured inclusive and prompt \Jpsi production in p$\bar{\rm p}$ collisions at $\sqrts=1.96$~TeV (Run II) over a rapidity of $|y| < 0.6$ for all transverse momenta from 0 to 20~$\gevc$~\cite{CDF05}. 
The inclusive \Jpsi cross section is: 
\newline
$\sigma_{\mpp} (|y|<0.6)= 4.08 \pm 0.02~{\rm (stat.)} ^{+0.36}_{-0.33}~{\rm (syst.)}~\mu{\rm b}$~\cite{CDF05}. 
\newline
The mean transverse momentum $\langle \pt \rangle$ was obtained from the data and is quoted in Table.~\ref{table1}.

\subsection{ALICE results}
\label{sec:dataalice}

The ALICE experiment at the LHC has studied inclusive \Jpsi production in two rapidity regions, $|y|<0.9$ and $-4.0<y<-2.5$, in \pp collisions at $\sqrts=7$~TeV~\cite{ALICEjpsi10}. 
The inclusive production cross sections are:  
\newline
$\dsigmady~|_{|y|<0.9}~=~5.97~\pm~1.02$~$\mu$b, and 
\newline
$\dsigmady~|_{-4.0<y<-2.5}~=~4.21~\pm~0.54$~$\mu$b. 
\newline
At forward-rapidity, ALICE \cite{ALICEjpsi10} has measured the differential cross section in six rapidity bins:
\newline
$\dsigmady~|_{-2.8<y<-2.5}~=~5.12~\pm~0.99$~$\mu$b,
\newline
$\dsigmady~|_{-3.1<y<-2.8}~=~4.34~\pm~0.62$~$\mu$b,
\newline
$\dsigmady~|_{-3.4<y<-3.1}~=~4.64~\pm~0.66$~$\mu$b,
\newline
$\dsigmady~|_{-3.7<y<-3.4}~=~3.59~\pm~0.53$~$\mu$b, and
\newline
$\dsigmady~|_{-4.0<y<-3.7}~=~3.05~\pm~0.54$~$\mu$b.
\newline
The $\langle \pt \rangle$ from both the mid- and forward-rapidity measurements are summarized in Table.~\ref{table1}.


\subsection{CMS results}
\label{sec:datacms}

The CMS experiment at the LHC has studied prompt and inclusive \Jpsi production in $|y|<2.4$ in \pp collisions at $\sqrts=7$~TeV~\cite{CMSjpsi10}. Only the measurement in the rapidity range $1.6<|y|<2.4$ has been considered in the present work, since in this rapidity range the CMS acceptance for \Jpsi goes down to $\pt \sim 0$. 
The $\pt$-integrated inclusive production cross section is: 
\newline
$\textrm{BR}_{ll} \times \dsigmady~|_{1.6<y<2.4} = 424 \pm 50$~nb. 
\newline
The mean transverse momentum (from data) is quoted in Table.~\ref{table1}.

\subsection{LHCb results}
\label{sec:datalhcb}

The LHCb experiment at the LHC has also measured inclusive and prompt \Jpsi production in the rapidity region of $2.0<y<4.5$ in pp collisions at $\sqrts=7$~TeV~\cite{LHCbjpsi10}. The inclusive production cross section is: 
\newline
$\sigma_{\mpp}(2.0<y<4.5)=~11.65~\pm~1.55$~$\mu$b.
\newline
The mean transverse momentum obtained from data is reported in Table.~\ref{table1}. 
The LHCb experiment measured the \Jpsi inclusive cross section for five rapidity bins in the rapidity range $2<y<4.5$ \cite{LHCbjpsi10}:
\newline
$\dsigmady~|_{2.0<y<2.5}~=~6.20~\pm~0.82$~$\mu$b, $\langle \pt \rangle_{2.0<y<2.5} = 2.58 \pm 0.19$~$\gevc$,
\newline
$\dsigmady~|_{2.5<y<3.0}~=~5.70~\pm~0.69$~$\mu$b, $\langle \pt \rangle_{2.5<y<3.0} = 2.59 \pm 0.09$~$\gevc$,
\newline
$\dsigmady~|_{3.0<y<3.5}~=~4.94~\pm~0.59$~$\mu$b, $\langle \pt \rangle_{3.0<y<3.5} = 2.51 \pm 0.09$~$\gevc$,
\newline
$\dsigmady~|_{3.5<y<4.0}~=~3.81~\pm~0.46$~$\mu$b, $\langle \pt \rangle_{3.5<y<4.0} = 2.42 \pm 0.09$~$\gevc$, and
\newline
$\dsigmady~|_{4.0<y<4.5}~=~2.64~\pm~0.33$~$\mu$b, $\langle \pt \rangle_{4.0<y<4.5} = 2.32 \pm 0.10$~$\gevc$.

\subsection{Other results not considered in this work}
\label{sec:dataother}

The \Jpsi production in \pp collisions was studied also by the STAR experiment at RHIC ($\sqrts=200$~GeV)~\cite{STAR} and ATLAS experiment at LHC ($\sqrts=7$~TeV)~\cite{ATLASjpsi10}. However, these measurements don't extend down to the low $\pt$ region and hence don't provide the $\pt$-integrated cross section for their respective rapidity range. Since these cross sections are the basic ingredients of the interpolation discussed in this work, these measurements were not considered. 
Results obtained by the D0 collaboration~\cite{D099} at Tevatron have been discarded here since they are given as a function of the \Jpsi pseudo-rapidity\footnote{
The \Jpsi rapidity for a given pseudo-rapidity depends on its transverse momentum: since the other results (and our calculation) are given in terms of the rapidity, we can not include this measurement in our work. 
}. 
SPS results~\cite{SPS} have been excluded since they refer to much lower interaction energies.

%

\section{Interpolation of \Jpsi cross section at mid-rapidity}
\label{sec:energyinterpolation}

The energy evolution of the inclusive \Jpsi production cross section in proton-proton and proton-antiproton collisions at mid-rapidity is shown in Fig.~\ref{fig:PolInterpolation}. Here we discuss three approaches to determine the $\pt$-integrated mid-rapidity cross section at $\sqrts=2.76$ and 5.5~TeV. The first method is a functional form fitting interpolation, while the other two are based on theoretical model predictions, FONLL~\cite{FONLL} and LO CEM~\cite{CEM}.

\subsection{A functional form fitting approach} 
\label{sec:polyfit}

\begin{figure}[!hbtp]
   \centering
   \includegraphics[width=0.7\textwidth]{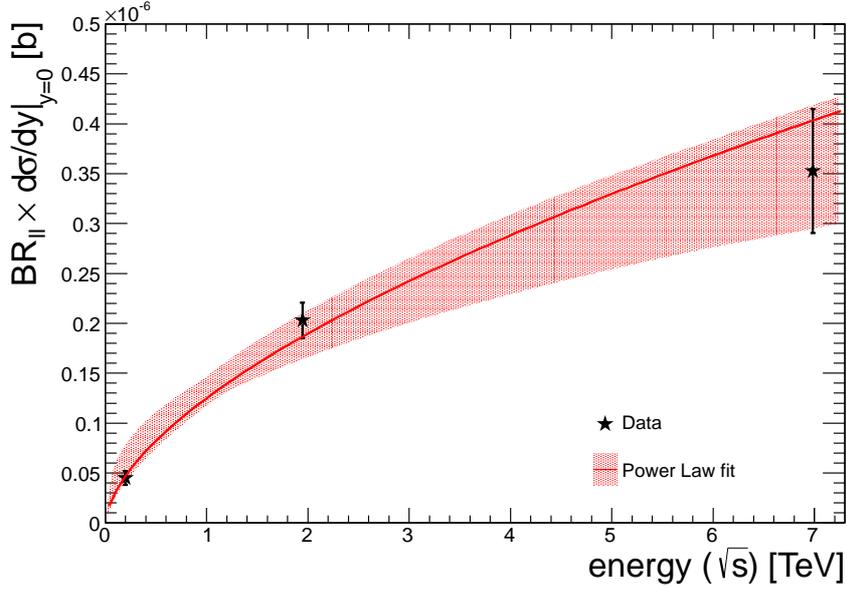}
   \caption{\label{fig:PolInterpolation} Energy dependence of inclusive \Jpsi production cross section at mid-rapidity as fitted by a power-law function: $BR_{ll} \times \dsigmady|_{y=0}(\sqrts)  = A \times \left(\sqrt{s/s_0}\right)^b$ where $A=124\pm9$ nb, $b=0.60\pm0.06$ and $\sqrt{s_0}=1$ TeV. }
\end{figure}

The cross section energy evolution can be described by a functional form, a power-law distribution:
\begin{equation}
f(\sqrts) = A\cdot \left(\sqrt{s/s_0}\right)^b,
\end{equation}
where $A$, $s_0$ and $b$ are free parameters. Although such characterization does not rely on any theoretical basis, it reproduces the measured distribution. The results are depicted in Fig.\ref{fig:PolInterpolation} and the interpolation results are reported in Table~\ref{tab:power_law_mid_rapidity}. 
The uncertainties on the cross section interpolation to $\sqrts=2.76$ and 7~TeV were estimated by re-doing the fit using the upper and lower limits of the PHENIX, CDF and ALICE uncertainties in every possible combination. The envelope of these results defines the uncertainty. 

\begin{table}[!htb]
\begin{center}
\begin{tabular}{||c|c||}
\hline \hline 
	$\sqrts$ 	&	 $BR_{ll} \times\sigma$ \\[0.5ex] \hline
	2.76 TeV	&	$230^{+22}_{-37}$ nb 	\\[0.5ex]
	5.5 TeV     &	$349^{+17}_{-83}$ nb	\\[0.5ex] \hline \hline
\end{tabular}
\end{center}
\caption{\label{tab:power_law_mid_rapidity} Mid-rapidity cross section results at 2.76~TeV and 5.5~TeV as interpolated by the power-law function.}
\end{table}

\subsection{FONLL based approach}
\label{subsec:fonllmidrap}

\begin{figure}[!htbp]
   \centering
   \includegraphics[width=0.75\textwidth]{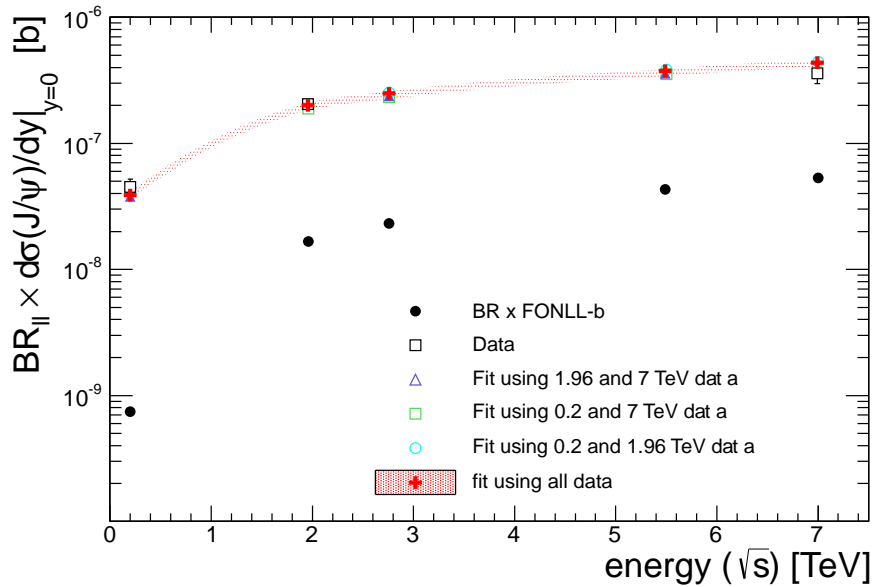}
   \caption{Energy dependence of inclusive \Jpsi production cross section as fitted by Eq.~\ref{eq:crsecFONLL} using charm and beauty FONLL calculations with their central parameterization.}
   \label{fig:FONLLInterpolation}
\end{figure}

Another approach consists to use the pQCD energy dependence of charm production cross section at mid-rapidity to describe the inclusive \Jpsi production cross section at mid-rapidity. The B feed-down contribution to the inclusive \Jpsi production cross section is considered via the pQCD beauty production cross section, weighted by the decay branching ratio, BR($B \longrightarrow J/\psi$)$\equiv BR$, as: 
\begin{equation}
\label{eq:crsecFONLL}
 \frac{{\rm d}\sigma}{{\rm d}y}\Bigr|_{y=0} = \alpha \times \frac{{\rm d}\sigma_{c}}{{\rm d}y}\Bigr|_{y=0}+2 \, BR \times \frac{{\rm d}\sigma_{b}}{{\rm d}y}\Bigr|_{y=0}. 
\end{equation}
Where $\alpha$ is a free parameter, $BR=(1.16 \pm 0.10) \times 10^{-2}$~\cite{pdg} and the charm and beauty differential cross sections are calculated in pQCD at Fixed Order with Next to Leading Log resummation FONLL~\cite{FONLL,cacciari}. 
The FONLL central predictions were obtained with the CTEQ6.6~\cite{CTEQ6.6} parton distribution functions (PDFs). 
The fit of the $\sqrts$ variation of \Jpsi inclusive production cross section with the function of Eq.~\ref{eq:crsecFONLL}, using the FONLL central parameters is shown in Fig.~\ref{fig:FONLLInterpolation}. 

The FONLL calculation uncertainties were evaluated by varying the PDFs, the quark masses, and the renormalization and factorization scales. These uncertainties are defined by the envelope of the obtained cross section, and are practically described by upper and lower curves. 
To account for the calculation uncertainties in the interpolation procedure, we refit the distribution considering these upper and lower curves and recalculate the interpolated cross sections. 
%
Finally, the interpolation is validated by observing how the results deviate when removing, one by one, the experimental points with the central FONLL parameterization. 
The results are reported in Table~\ref{tab:SigmaMidFONLL}. 
\begin{table}[!htbp]
\begin{center}
\begin{tabular}{||c|cccccc||} \hline \hline
Fit      & Data         & \multicolumn{5}{c||}{$BR_{ll} \times \dsigmady|_{y=0}$~[nb] per $\sqrts$~[TeV] } \\ 
 FONLL & excluded  &         0.2 	&      1.96 	&    2.76 	&       5.5 	&        7 \\  \hline
central & none 		&{\it39 $\pm$ 3}& {\it196 $\pm$ 14}&  241 $\pm$ 17	&  366 $\pm$ 25	& {\it422 $\pm$ 28} \\
lower   & none 		&{\it45 $\pm$ 3}& {\it201 $\pm$ 14}&  229 $\pm$ 16	&  299 $\pm$ 20	& {\it332 $\pm$ 21} \\
upper   & none 		&{\it41 $\pm$ 3}& {\it195 $\pm$ 14}&  238 $\pm$ 17	&  357 $\pm$ 24	& {\it409 $\pm$ 27} \\
central  & RHIC 	&    38 $\pm$ 3 & {\it190 $\pm$ 15}&  235 $\pm$ 19	&  356 $\pm$ 28	& {\it411 $\pm$ 31} \\ 
central & Tevatron 	&{\it37 $\pm$ 5}&     189 $\pm$ 22 &  232 $\pm$ 27	&  353 $\pm$ 40	&{\it407 $\pm$ 45} \\
central  & LHC 		&{\it41 $\pm$ 3}& {\it205 $\pm$ 16}&  253 $\pm$ 19	&  383 $\pm$ 28	&    442 $\pm$ 32 \\
\hline \hline
\end{tabular}
\end{center}
\caption{Inclusive \Jpsi production cross sections at mid-rapidity obtained with the FONLL based approach. In order to distinguish the interpolated values from the other fit values at a given energy, the latter are quoted in italics. }
\label{tab:SigmaMidFONLL}
\end{table}
The cross sections evaluated when removing an experimental point to the given energy are in fair agreement (within uncertainties) with the measurements (see Sec.~\ref{data}). 
The extrapolated values obtained with the three FONLL curves are compatible, 
although we note that the fit with the lower FONLL uncertainty band tends to give significantly lower results than the others at high energies
\footnote{
This effect is understood as related to the discontinuous shape of the FONLL lower-band uncertainties for direct charm production~\cite{FONLL,cacciari}.
}, this is a source of systematic uncertainty. 
We remark that the results of the FONLL interpolation are in good agreement with those of the functional form fit presented in the previous subsection (Sec.~\ref{sec:polyfit}).

\subsection{LO CEM based interpolation}
\label{subsec:cemmidrap}

\begin{figure}[!hbtp]
   \centering
   \includegraphics[width=0.75\textwidth]{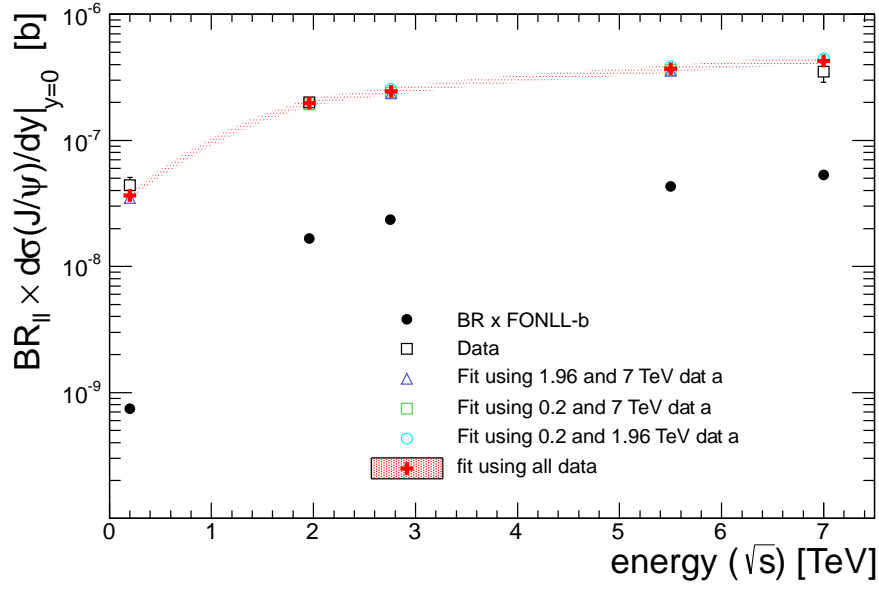} 
   \caption{Energy dependence of inclusive \Jpsi production cross section as fitted by Eq.~\ref{eq:crsecCEM}, with the \Jpsi LO CEM and the beauty FONLL calculations, and the CTEQ6 parton distribution functions.}
   \label{fig:CEMInterpolation}
\end{figure}

The third approach uses the prompt \Jpsi cross section energy dependence as computed in the Colour Evaporation Model (CEM) at Leading Order (LO)~\cite{CEM}. As in the previous subsection (Sec.~\ref{subsec:fonllmidrap}), the B-decay contribution is estimated via the FONLL beauty predictions. The fit is defined such that the normalization of the LO CEM calculation is left as a free parameter, $\alpha$: 
\begin{equation}
\label{eq:crsecCEM}
\frac{{\rm d}\sigma}{{\rm d}y}\Bigr|_{y=0} = \alpha \times \frac{{\rm d}\sigma_{CEM}}{{\rm d}y}\Bigr|_{y=0} + 2 \, BR \times \frac{{\rm d}\sigma_{b}}{{\rm d}y}\Bigr|_{y=0}. 
\end{equation}
Different LO PDFs have been considered for this exercise\footnote{
It is worth noting that LO CEM based interpolation results using other parton distribution functions are compatible within uncertainties with the ones reported here.
}. Here, we only report the results obtained with CTEQ6~\cite{CTEQ6} and MRST01~\cite{MRST01}. 
The best fit is obtained with the CTEQ6 PDF set and is shown in Fig.~\ref{fig:CEMInterpolation}. 
Similarly to the other approaches, the interpolation has been validated by removing the data points one by one from the fit. 
The results are summarized in Table~\ref{tab:SigmaMidCEM}. 
\begin{table}[!hbtp]
\begin{center}
\begin{tabular}{||c|cccccc||} \hline \hline
Fit      & Data         & \multicolumn{5}{c||}{$BR_{ll} \times  \dsigmady|_{y=0}$~[nb] per $\sqrts$ [TeV] } \\ 
LO CEM & excluded & 0.2 	 &        1.96     &       2.76         &      5.5 	&         7 \\  \hline
CTEQ6 & none     & {\it36 $\pm$ 3}&{\it198 $\pm$ 14}&  245 $\pm$ 17	&  370 $\pm$ 25	&{\it425 $\pm$ 29}\\ 
MRST01& none    & {\it29 $\pm$ 2}&{\it194 $\pm$ 14}&  248 $\pm$ 18	&  399 $\pm$ 28	&{\it470 $\pm$ 33}\\ 
CTEQ6 &  RHIC    &    35 $\pm$ 3 &{\it191 $\pm$ 15}&  236 $\pm$ 19	&  356 $\pm$ 28	&{\it410 $\pm$ 31}\\
CTEQ6 & Tevatron & {\it36 $\pm$ 5}&    193 $\pm$ 23 &  239 $\pm$ 28	&  362 $\pm$ 41	&{\it416 $\pm$ 47}\\
CTEQ6 & LHC      & {\it38 $\pm$ 3}&{\it208 $\pm$ 16}&  257 $\pm$ 19	&  388 $\pm$ 28	&    446 $\pm$ 32 \\ 
\hline \hline
\end{tabular}
\end{center}
\caption{Inclusive \Jpsi production cross sections at mid-rapidity obtained with the LO CEM based interpolation. In order to  distinguish the interpolated values from the other fit values at a given energy, the latter are quoted in italics.}
\label{tab:SigmaMidCEM}
\end{table}
The CTEQ6 and MRST01 derived calculations are consistent with each other. 
The cross sections obtained when removing experimental points are also in good agreement (within uncertainties) with the measurements (see Sec.~\ref{data}). 
These results also agree with those of the two previous approaches (Sec.~\ref{sec:polyfit} and Sec.~\ref{subsec:fonllmidrap}).

\subsection{Results}

The values obtained for the inclusive \Jpsi cross sections at $\sqrts=2.76$ and 5.5~TeV, at mid-rapidity, using the functional form, the FONLL and the LO CEM based fits are presented in Table~\ref{tab:final_midrap}, together with the $\chi ^2$ and the number of degrees of freedom of the fit. 
\begin{table}[!htbp]
\begin{center}
\begin{tabular}{||l|cc||} \hline \hline
  Fit  &$\alpha$& $\chi^2$ (ndf) \\  \hline
 $A \times \left(\sqrt{s/s_0}\right)^b$&                &  1.5 (1)  \\  
   FONLL central   & $(11.4 \pm 0.9) \cdot 10^{-3}$ &  1.0 (2) \\  
   FONLL upper     &  $(4.9 \pm 0.4) \cdot 10^{-3}$&  0.57 (2)\\  
   FONLL lower     &  $(2.9 \pm 0.2) \cdot 10^{-2}$ &  0.06 (2) \\ 
   LO CEM CTEQ6    &  $1.94 \pm 0.15$   &  1.3 (2) \\  
   LO CEM MRST01   &  $4.7 \pm 0.4$     &  4.4 (2) \\  \hline \hline
  Fit   & \multicolumn{2}{c||}{$BR_{ll} \times \dsigmady|_{y=0}$~[nb] }  \\  
                         &       $\sqrts=2.76$ TeV    &     $\sqrts= 5.5$ TeV        \\ [0.5ex] \hline 
 $A \times \left(\sqrt{s/s_0}\right)^b$& $230 ^{+23}_{-37}$& $349 ^{+18}_{-83}$  \\[0.5ex]
   FONLL central   &  $241 \pm 17$  & $366 \pm 25$      \\  
   FONLL upper     &  $238 \pm 17$  & $357 \pm 24$  \\  
   FONLL lower     &  $229 \pm 16$  & $299 \pm 20$     \\ 
   LO CEM CTEQ6    &  $245 \pm 17$  & $370 \pm 25$    \\  
   LO CEM MRST01   &  $248 \pm 18$  & $399 \pm 28$   \\[0.5ex] \hline
   Result for inclusive \Jpsi & $239^{+6}_{-10}~\textrm{(model)} \pm 31~\textrm{(fit)}$ & 
   	$350^{+20}_{-51}~\textrm{(model)} \pm 51~\textrm{(fit)}$ \\[0.5ex]
   Estimate for prompt \Jpsi &  $215^{+9}_{-11}~\textrm{(model)} \pm 28~\textrm{(fit)}$ & 
   	 $315^{+21}_{-47}~\textrm{(model)} \pm 46~\textrm{(fit)}$ \\[0.5ex]
   \hline  \hline
\end{tabular}
\end{center}
\caption{Extrapolated values of the inclusive \Jpsi production cross section at mid-rapidity according to the phenomenological, FONLL and LO CEM fits. The fit $\chi^2$ and the number of degrees of freedom (ndf) are also reported.  
}
\label{tab:final_midrap}
\end{table}
The fit uncertainties at $\sqrts=2.76$~(5.5)~TeV are of $7\%$ for the LO CEM and the FONLL fits, while they are of $13\%$ ($15\%$) for the functional form approach\footnote{
When dealing with asymmetric uncertainties, we quote as relative error the average between the upper and the lower value of the asymmetric uncertainties.
}. To combine these results we: 
\begin{enumerate}[i)]
\item discard the LO CEM MRST01 fit, that has a $\chi ^2$/ndf $>2$, 
\item compute the average of the three FONLL values, 
\item estimate the weighted average of the  power-law fit, the FONLL-average (ii) and the LO CEM CTEQ6 results.
\end{enumerate} 
We sort the uncertainties of the combined result in two components. One is related to the spread of the values obtained with the different approaches, resulting in an uncertainty band covering the whole range of values. We will refer to this component of the uncertainty as the {\sf model} related one.
The other component is related to the fit uncertainties and includes the influence of the experimental measurements uncertainties. For this contribution, that we will quote as {\sf fit} uncertainty, we (conservatively) assume the largest of the relative fit uncertainties (the functional form fit one), that is of 13\% (15\%) at 2.76~(5.5)~TeV. 
The combined results for the inclusive \Jpsi cross section at mid-rapidity are quoted in Table~\ref{tab:final_midrap}.

Finally, an estimate of the prompt \Jpsi production cross section at mid-rapidity can be reported subtracting the B feed-down contribution. 
The LHCb results at $\sqrts=7$~TeV and $2.0<y<4.5$ indicate that the contribution from B-decays is 10\% with a 15\% relative uncertainty~\cite{LHCbjpsi10}. 
CDF~\cite{CDF05} measured the B feed-down fraction as a function of $\pt$ for $\pt>1.25$~GeV/c. If we consider that for $\pt<1.25$~GeV/c it is $\sim 9\%$, we obtain a $\pt$-integrated B-decay contribution of 10.5\%. 
Assuming that the B-decay contribution to the inclusive \Jpsi production cross section at 2.76 and 5.5 TeV is 10\%  with a (conservative) uncertainty of 30\%, we obtain the prompt \Jpsi cross section at mid-rapidity reported in Table~\ref{tab:final_midrap}. 

%

\section{Rapidity distribution}
\label{sec:rapidityinterpolation}

The knowledge of the rapidity dependence of \Jpsi production at 2.76 and 5.5 TeV is crucial in order to provide a reference for the measurements performed at large rapidities. Here we focus on the estimates at $y = 3.25$ (ALICE muon spectrometer) and $y = 2$ (region where the CMS acceptance for quarkonia goes down to low $\pt$). 
We consider and compare the interpolation results based on the FONLL and the LO CEM calculations used for the calculation of the mid-rapidity cross section (Sec.~\ref{sec:energyinterpolation}) with those obtained with a functional form fit approach. 
For this exercise, we consider the measurements at RHIC~\cite{PHENIX07} and at the LHC~\cite{ALICEjpsi10,CMSjpsi10,LHCbjpsi10}.


\subsection{Interpolation based on theoretical calculations}
\label{subsec:modelsInterpolation}

\begin{figure}[!htbp]
   \centering
   \includegraphics[width=0.75\textwidth]{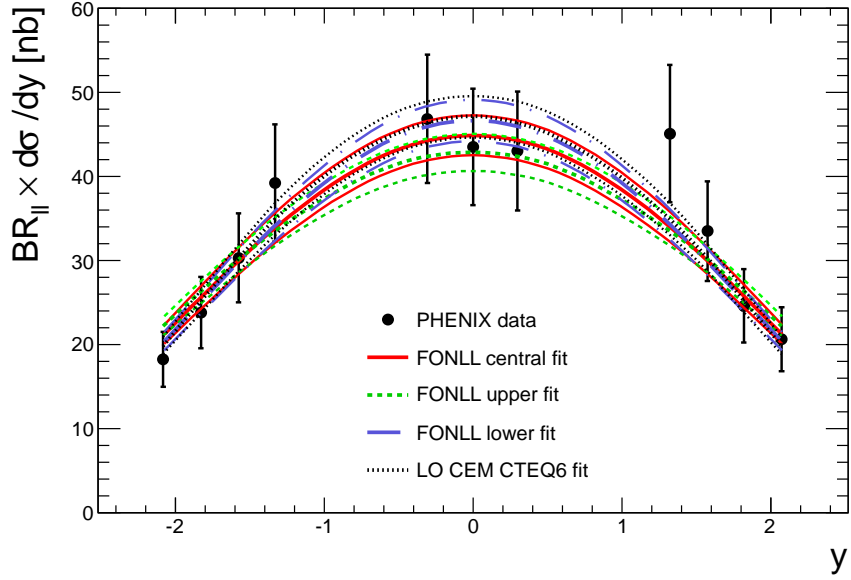} 
   \caption{\label{fig:RapFitModel200GeV}
   Inclusive \Jpsi rapidity distribution at $\sqrts=200$~GeV~\cite{PHENIX07} as fitted using the LO CEM and the FONLL calculations. The measurement uncertainties are obtained by summing in quadrature the PHENIX systematic and statistical uncertainties.}
\end{figure}

\begin{figure}[!htbp]
   \centering
   \includegraphics[width=0.75\textwidth]{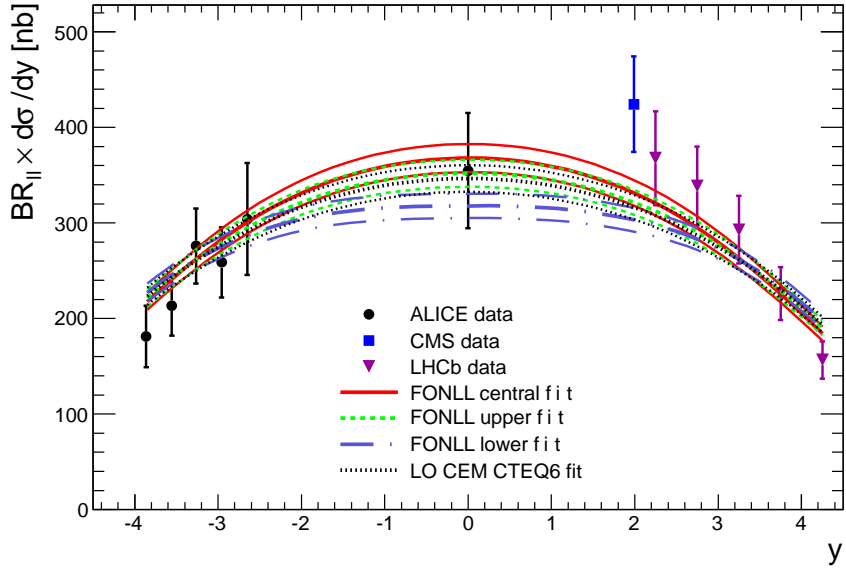}
   \caption{\label{fig:RapFitModel7TeV}
   Inclusive \Jpsi rapidity distribution at $\sqrts=7$~TeV~\cite{ALICEjpsi10,CMSjpsi10,LHCbjpsi10} as fitted using the LO CEM and the FONLL calculations. The data uncertainties are obtained by summing in quadrature the measurement systematic and statistical uncertainties.}
\end{figure}

The FONLL and LO CEM calculations (see Sec.~\ref{subsec:fonllmidrap} and Sec.~\ref{subsec:cemmidrap}) provide predictions for the production cross section rapidity dependence. 
In order to test their ability to predict the inclusive \Jpsi rapidity shape at different energies, we fit the measured rapidity distributions at $\sqrts=200$~GeV and 7~TeV with a model-derived function. 
The chosen fit function is similar to the $\sqrts$ form of Eq.~\ref{eq:crsecFONLL} and Eq.~\ref{eq:crsecCEM} and is defined as:
\begin{equation}
\label{eq:ycrsec}
\frac{{\rm d}\sigma}{{\rm d}y} = \beta \, \left[ \alpha_{\rm model} \times \frac{{\rm d}\sigma_{\rm model}}{{\rm d}y} +2 \, BR \times \frac{{\rm d}\sigma_{b}}{{\rm d}y} \right]. 
\end{equation}
We fix the (model dependent) $\alpha_{\rm model}$ parameters to be those obtained from the mid-rapidity fits~\footnote{
Fixing the $\alpha_{\rm model}$ parameter allows us to fix the relative normalization of the prompt and feed-down contributions at mid-rapidity.
} in Sec.\ref{sec:energyinterpolation}  and we let the overall normalization ($\beta$) be the only free parameter of the fit. 
The fits are shown in Figs.~\ref{fig:RapFitModel200GeV} and \ref{fig:RapFitModel7TeV}, and their $\chi^2$ values are summarized in Table~\ref{tab:test_rap}. 
We observe that \hbox{$\chi^2/{\rm ndf}<2$} for all the fits: all the models are compatible with the measured inclusive \Jpsi rapidity distribution at both 200~GeV and 7~TeV. The predictions of these models are expressed here in terms of an extrapolation factor from mid- to forward rapidity, defined as: 
\begin{equation}
\label{eq:FforwardDef}
f_{\rm forw}(y) =  \frac{{\rm d}\sigma}{{\rm d}y} (y)   \Big/   \frac{{\rm d}\sigma}{{\rm d}y} \Bigr|_{y=0}.
\end{equation} 
The values of $f_{\rm forw}$ for the considered models at the energies and rapidities of interest are reported in Table~\ref{tab:f_forw}. 
\begin{table}[!htbp]
\begin{center}
\begin{tabular}{||l|cccc||} \hline \hline
	      &\multicolumn{2}{c}{$\sqrts=200$~GeV} & \multicolumn{2}{c||}{$\sqrts=7$~TeV} \\
  Fit                   & $\chi^2$~(ndf)  & $\beta$       &  $\chi^2$~(ndf)  & $\beta$   \\  \hline
   FONLL central  & 4.1 (10)        & 1.10$\pm$0.06 & 10.9 (11)        & 0.87$\pm$0.03\\  
   FONLL upper    & 5.6 (10)        & 0.98$\pm$0.05 & 13.6 (11)        & 0.85$\pm$0.03\\  
   FONLL lower    & 3.6 (10)        & 0.98$\pm$0.05 & 19.0 (11)        & 0.95$\pm$0.04\\  
   LO CEM CTEQ6   & 3.1 (10)        & 1.29$\pm$0.07 & 15.2 (11)        & 0.82$\pm$0.03\\  
\hline \hline
     	& \multicolumn{4}{c||}{$f_{\rm forw} (y)$} \\
  Fit        &  \multicolumn{2}{c}{$\sqrts=2.76$~TeV}  & \multicolumn{2}{c||}{$\sqrts=5.5$~TeV} \\  
                        &  $y=2.0$  & $y=3.25$ & $y=2.0$ & $y=3.25$ \\  \hline
   FONLL central  &   0.85  & 0.59 & 0.88 & 0.68 \\  
   FONLL upper    &   0.87  & 0.63 & 0.90 & 0.72 \\  
   FONLL lower    &   0.91  & 0.69 & 0.94 & 0.80 \\ 
   LO CEM CTEQ6   &   0.87  & 0.64 & 0.90 & 0.73 \\[0.5ex] 
   Average & $0.87_{-0.03}^{+0.04}$     & $0.64_{-0.05}^{+0.05}$        & $0.90_{-0.02}^{+0.04}$        & $0.73_{-0.05}^{+0.07}$ \\[0.5ex]
     \hline\hline
     & \multicolumn{4}{c||}{$BR_{ll} \times \left. \dsigmady \right|^{\rm incl.}_{y=y_0}$ [nb]} \\[0.5ex]
     $y_0$ & \multicolumn{2}{c}{$\sqrts=2.76$~TeV}  & \multicolumn{2}{c||}{$\sqrts=5.5$~TeV} \\[0.5ex]  \hline
     2.0 &  \multicolumn{2}{c}{$208 \pm 28~\textrm{\scriptsize (corr.)} \, _{-6}^{+9}~\textrm{\scriptsize (uncorr.)}$} 
     	&  \multicolumn{2}{c||}{$316 \pm 65~\textrm{\scriptsize (corr.)} \, _{-7}^{+14}~\textrm{\scriptsize (uncorr.)}$} \\[0.5ex]
     3.25 &  \multicolumn{2}{c}{$153 \pm 21~\textrm{\scriptsize (corr.)} \, _{-12}^{+12}~\textrm{\scriptsize (uncorr.)}$} 
     	&  \multicolumn{2}{c||}{$256 \pm 53~\textrm{\scriptsize (corr.)} \, _{-16}^{+23}~\textrm{\scriptsize (uncorr.)}$} \\[0.5ex] \hline \hline
\end{tabular}
\end{center}
\caption{Ratios of the mid- to forward-rapidity \Jpsi inclusive cross section scaling as obtained with FONLL and LO CEM.
	The values of the fits $\chi^2$, the number of degrees of freedom (ndf), and the $\beta$ parameters at $\sqrts=$200~GeV and 7~TeV are reported. 
	The interpolated y-differential production cross sections at $\sqrts=2.76$ and 5.5~TeV are also quoted.}
\label{tab:f_forw}
\label{tab:test_rap}
\end{table}
The results are a combination of these values. We first compute the average of the FONLL estimates and then the average between the FONLL-average and the LO CEM CTEQ6 value. The uncertainties are defined by the envelope of all the values. 
The inclusive forward rapidity cross sections can then be obtained by multiplying the mid-rapidity cross sections of Table~\ref{tab:final_midrap} by the forward rapidity factors. 
The uncertainties are split in a correlated (from the mid-rapidity cross section) and an uncorrelated (from the $f_{\rm forw}$ factor) component. 
The results for the inclusive J/$\psi$ cross section at forward rapidities are reported on Table~\ref{tab:test_rap}. 

\begin{figure}[!hbtp]
   \centering
  \subfigure[$\sqrts=2.76$~TeV]{
           \includegraphics[width=0.625\textwidth,height=0.5\textwidth]{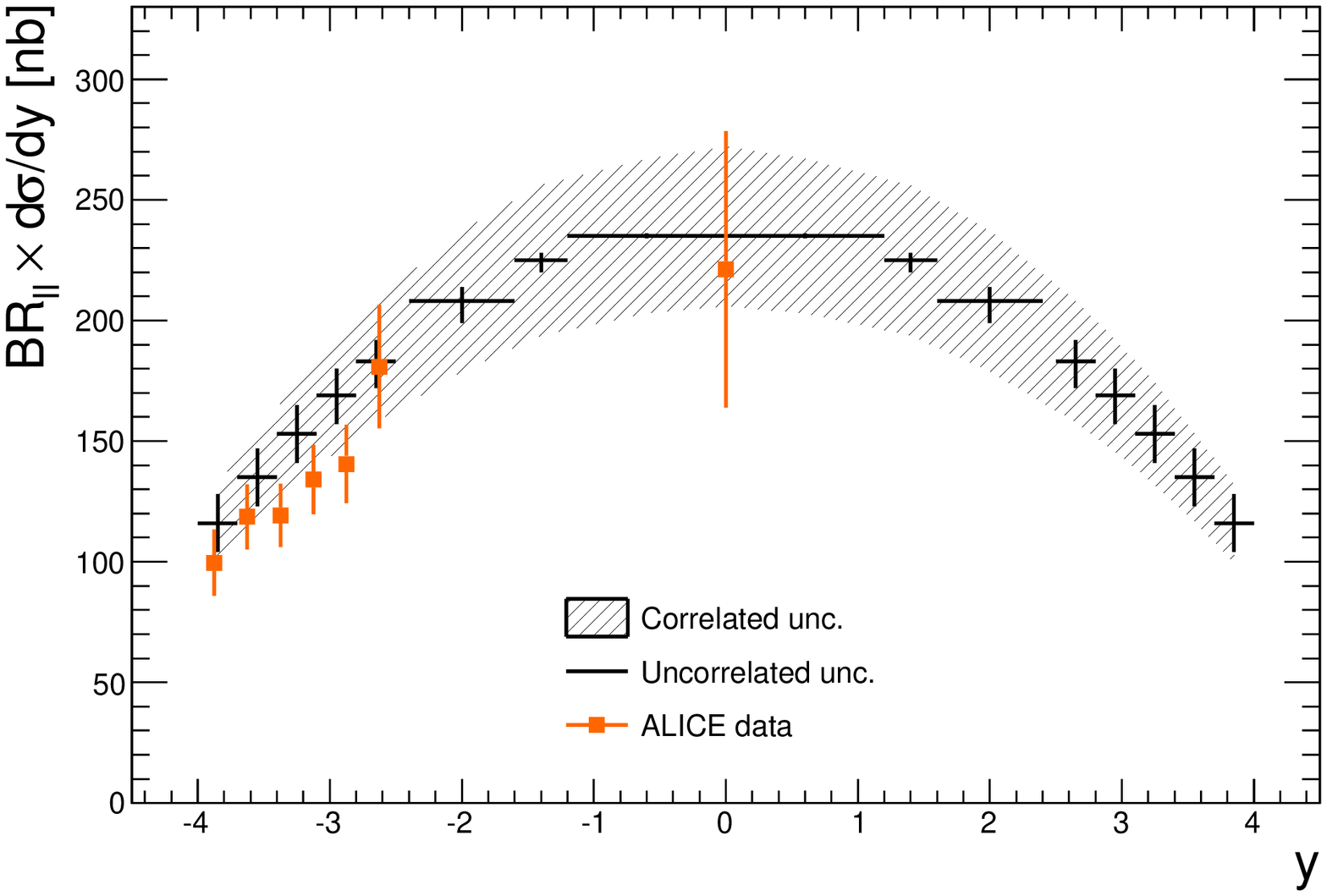}
           \label{fig:rapidityJpsi276}
           }
  \subfigure[$\sqrts=5.5$~TeV]{
           \includegraphics[width=0.625\textwidth,height=0.5\textwidth]{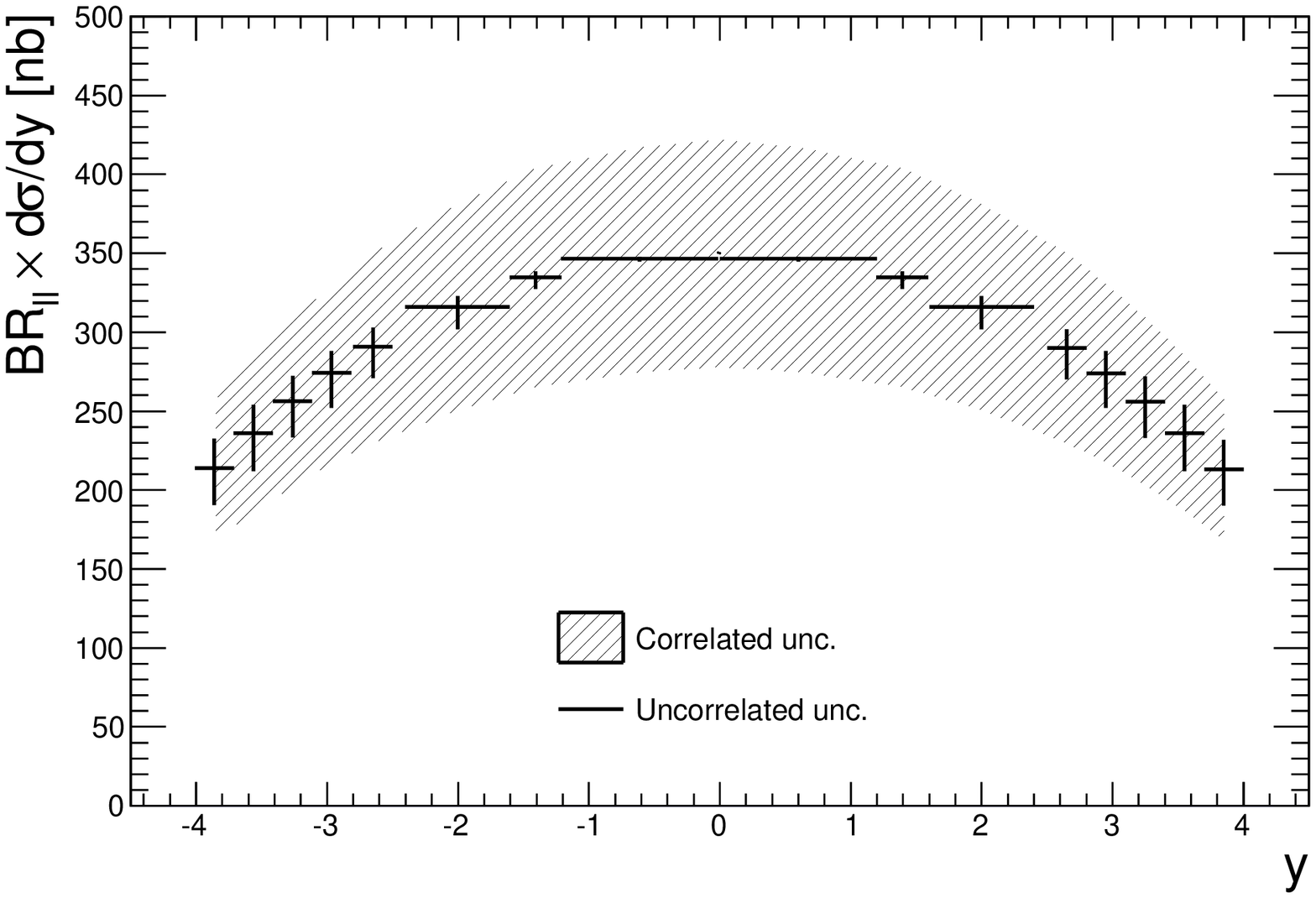}
           \label{fig:rapidityJpsi55}
           }
   \caption{Inclusive \Jpsi production rapidity distribution at $\sqrts=$2.76 TeV~\subref{fig:rapidityJpsi276} and $\sqrts=5.5$~TeV~\subref{fig:rapidityJpsi55}. The ALICE measurement at $\sqrts=$2.76 TeV are also shown with the total (statistical and systematical) uncertainties.
        }
\end{figure}

Similarly, we can calculate the expected rapidity-differential cross section in a number of rapidity bins of interest for the LHC experiments. The resulting rapidity distributions at 2.76 and 5.5 TeV are shown in Figs.~\ref{fig:rapidityJpsi276} and \ref{fig:rapidityJpsi55}.
  
\subsection{A functional form fit approach: universal rapidity distribution}
\label{subsec:urp}

\begin{figure}[!hbtp]
   \centering
  \subfigure[versus $y$]{
           \includegraphics[width=0.55\textwidth]{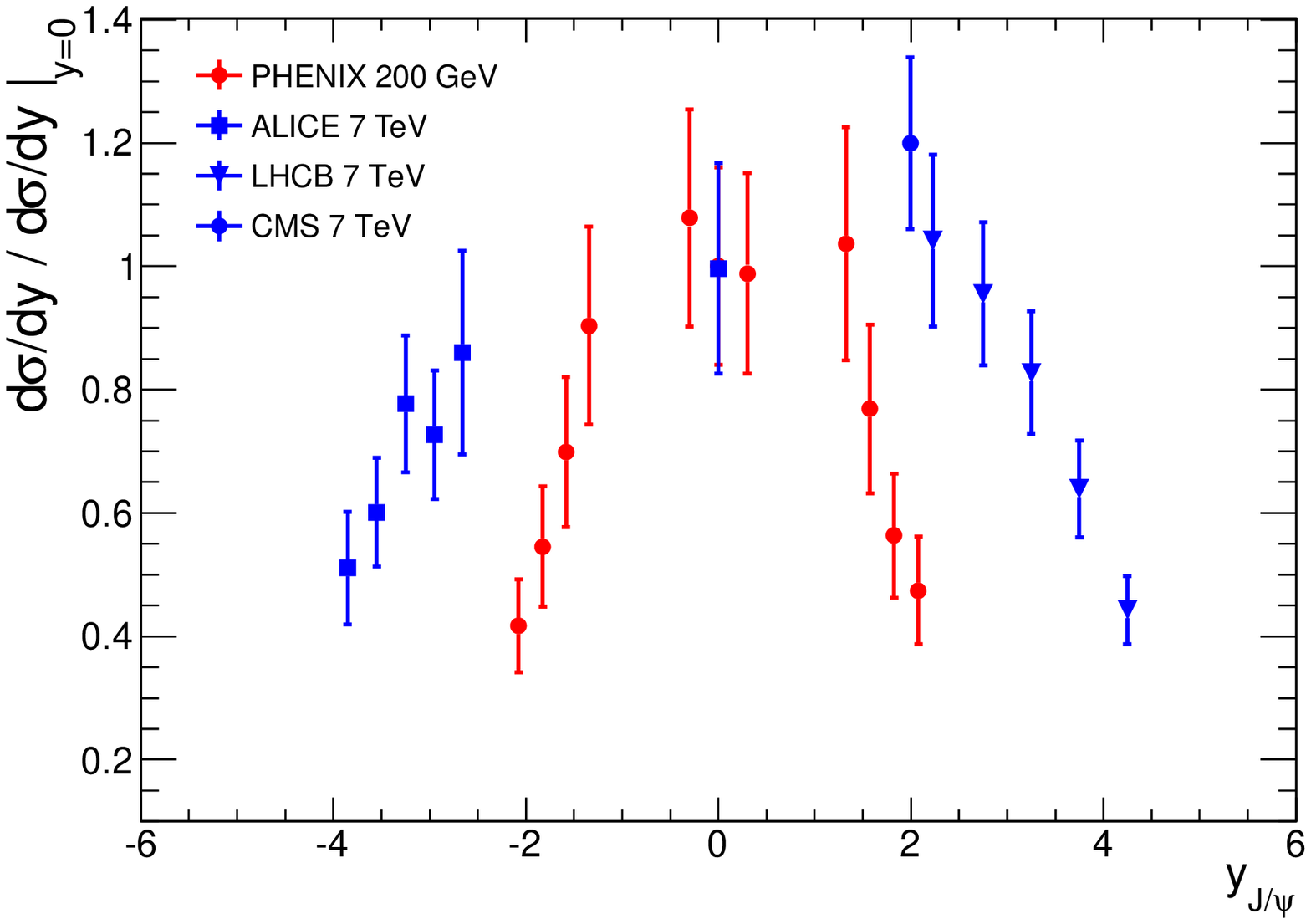}
           \label{fig:UniversalYjpsia}
           }
  \subfigure[versus $y/y_{\rm beam \, max}$]{
           \includegraphics[width=0.55\textwidth]{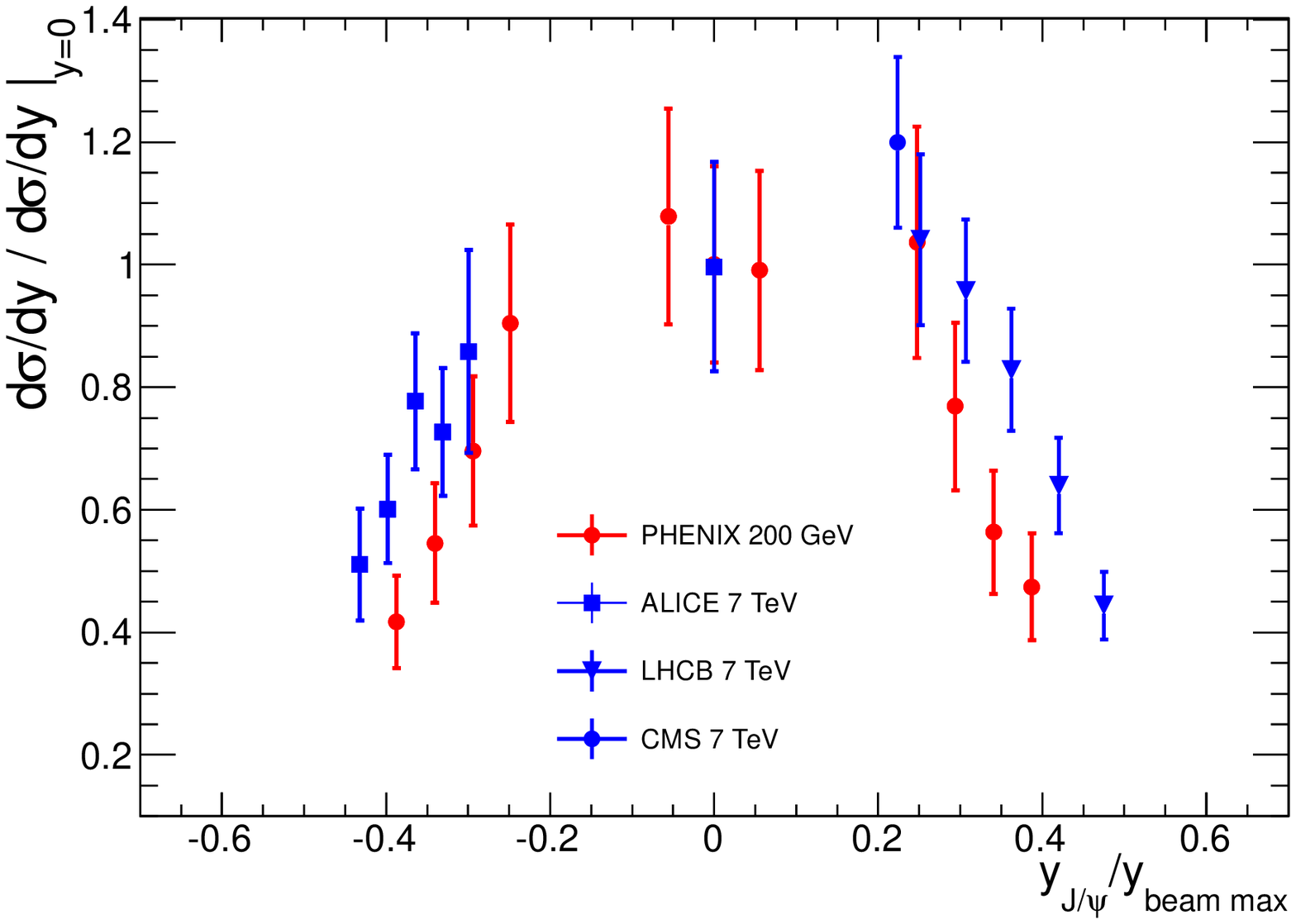}
           \label{fig:UniversalYjpsib}
           }
  \subfigure[versus $y/y_{\rm max}$]{
           \includegraphics[width=0.55\textwidth]{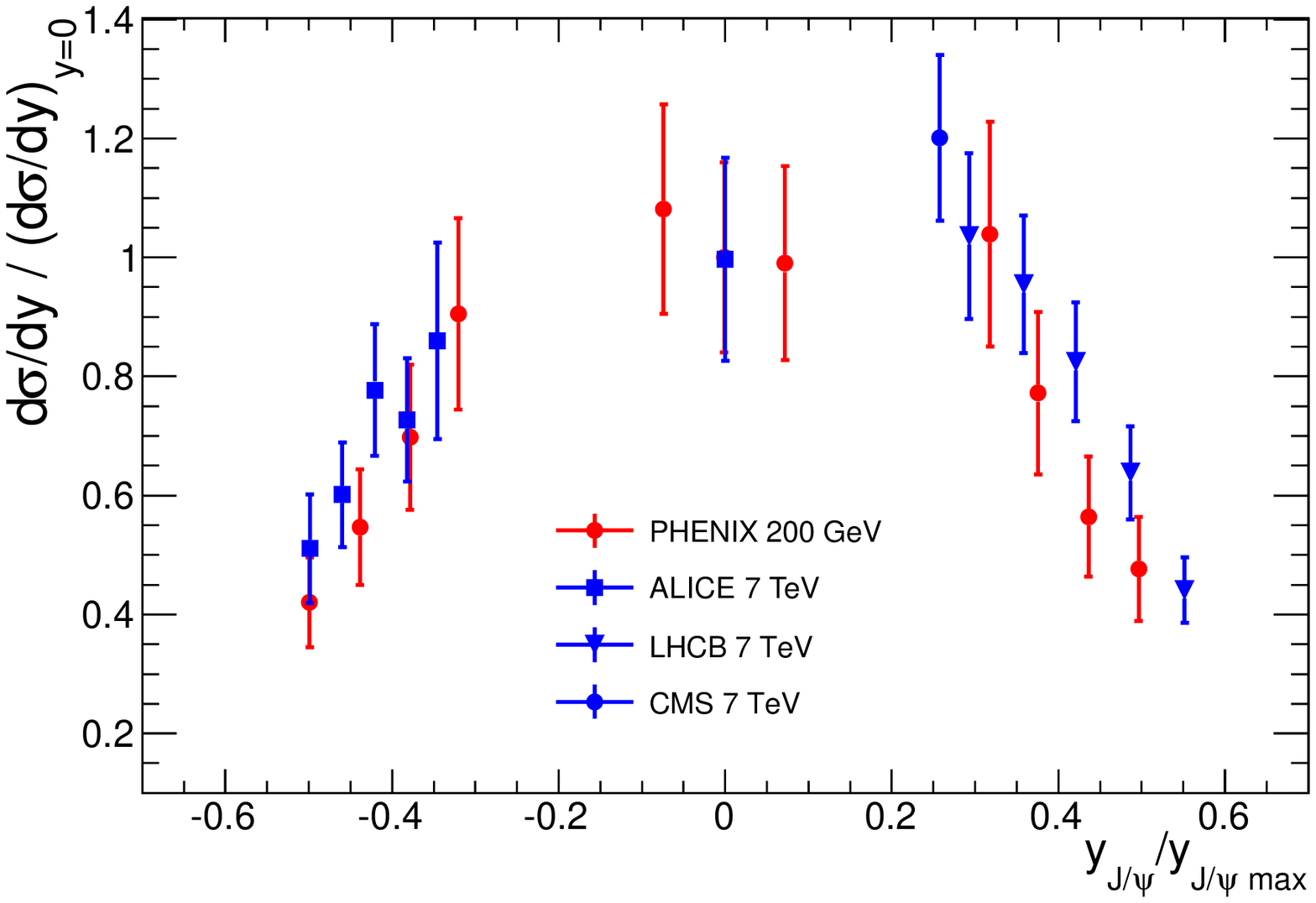}
           \label{fig:UniversalYjpsic}
           }
   \caption{\label{fig:UniversalYjpsi} Inclusive \Jpsi production cross section as a function of the \Jpsi rapidity~\subref{fig:UniversalYjpsia}, $y /y_{\rm beam \, max}$~\subref{fig:UniversalYjpsib}, $y /y_{\rm max}$~\subref{fig:UniversalYjpsic} in proton-proton collisions at 200 GeV (PHENIX~\cite{PHENIX07}) and at 7 TeV (ALICE, CMS and LHCb experiments~\cite{ALICEjpsi10, CMSjpsi10,LHCbjpsi10}).
        }
\end{figure}

In this subsection we explore a model independent approach to the extrapolation of the rapidity distribution, based on the search for an universal energy scaling behaviour in the rapidity distributions measured at different energies. The advantage of this approach is that there is no {\it a priori} assumption, and the results only depend on the existing measurements and their uncertainties. 
To compare the $y$-differential cross sections at 200~GeV and at 7~TeV, these values are renormalized by the $\pt$-integrated cross section at $y=0$. We draw the obtained values versus: $y$, $y/y_{\rm max}$ and $y/y_{\rm beam \, max}$, where $y$ is the \Jpsi rapidity, $y_{\rm max}=\ln{(\sqrts/m_{{\rm J}\psi} ) }$ and $y_{\rm beam \, max}=\ln{(\sqrts/m_{p} ) }$. The resulting distributions are shown in Fig.~\ref{fig:UniversalYjpsi}. 
The similarity between the distributions obtained at 200~GeV and 7~TeV is remarkable, the most striking case being the scaling with $y /y_{max}$. 
We then choose $y /y_{\rm max}$ as a scaling variable, and perform global fits to the RHIC and LHC data with suitable functions, the purpose being to define $\sqrts$-independent rapidity distribution functions that can be used for the calculations at $\sqrts$=2.76 and 5.5~TeV. 
We note that the values used to normalize the $y$-differential cross sections in Fig.~\ref{fig:UniversalYjpsi}, their value at y=0, is not an optimal choice since depends strongly on a single data point. In the following we will re-evaluate this normalization for each functional form. 

\begin{figure}
   \centering
   \includegraphics[width=0.8\textwidth]{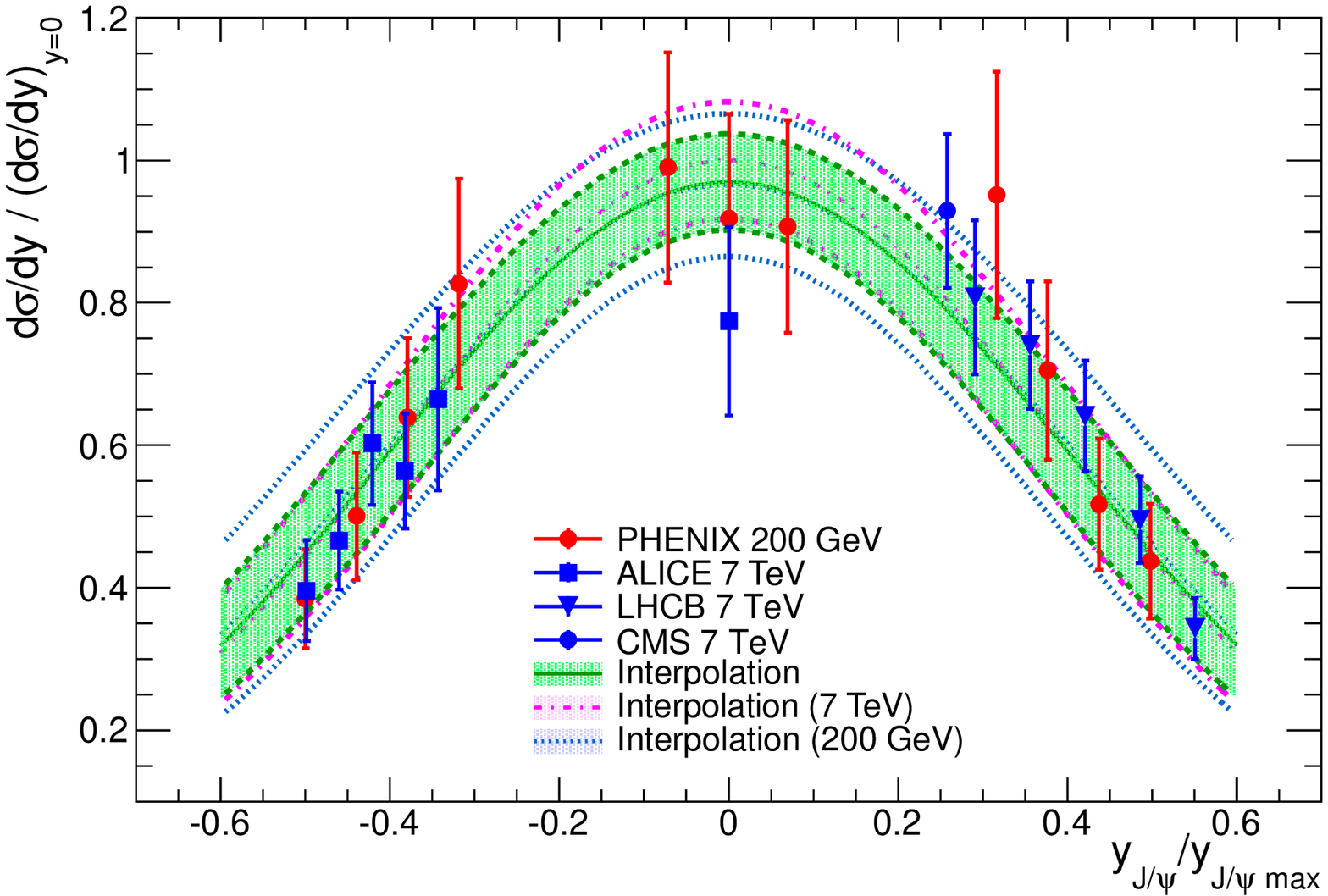}
   \caption{Normalised \Jpsi production cross section as a function of the relative rapidity $y /y_{\rm max}$. Three gaussian fits are shown: one was performed by using all data drawn here, other using only RHIC data and the last considering only the LHC data.}
   \label{fig:UniversalYreljpsi_Gaussian}
\end{figure}

The first fit function considered is a Gaussian distribution, we consider that: 
\begin{equation}
 \frac{{\rm d}\sigma}{{\rm d}y} \, \,/  \,\, \frac{{\rm d}\sigma}{{\rm d}y} \Bigr|_{y=0} =  e^{ -(y / y_{\rm max})^2/2\sigma_y^2}
\end{equation}
where $\sigma_y$ is the width of the Gaussian and $\left. {\rm d} \sigma / {\rm d}y \right|_{y=0}$ is a normalization factor obtained via Gaussian fit separately for the two energies. The result of the fit is shown in Fig.~\ref{fig:UniversalYreljpsi_Gaussian}. 
For comparison, the RHIC-only and the LHC-only fit results are also drawn, showing that the fit is stable versus excluding data points at a given energy. Also, it is worth noting that uncertainties are reduced when considering all data points together. 

\begin{figure}[!hbtp]
   \centering
  \subfigure[$n=2$]{
           \includegraphics[width=0.47\textwidth]{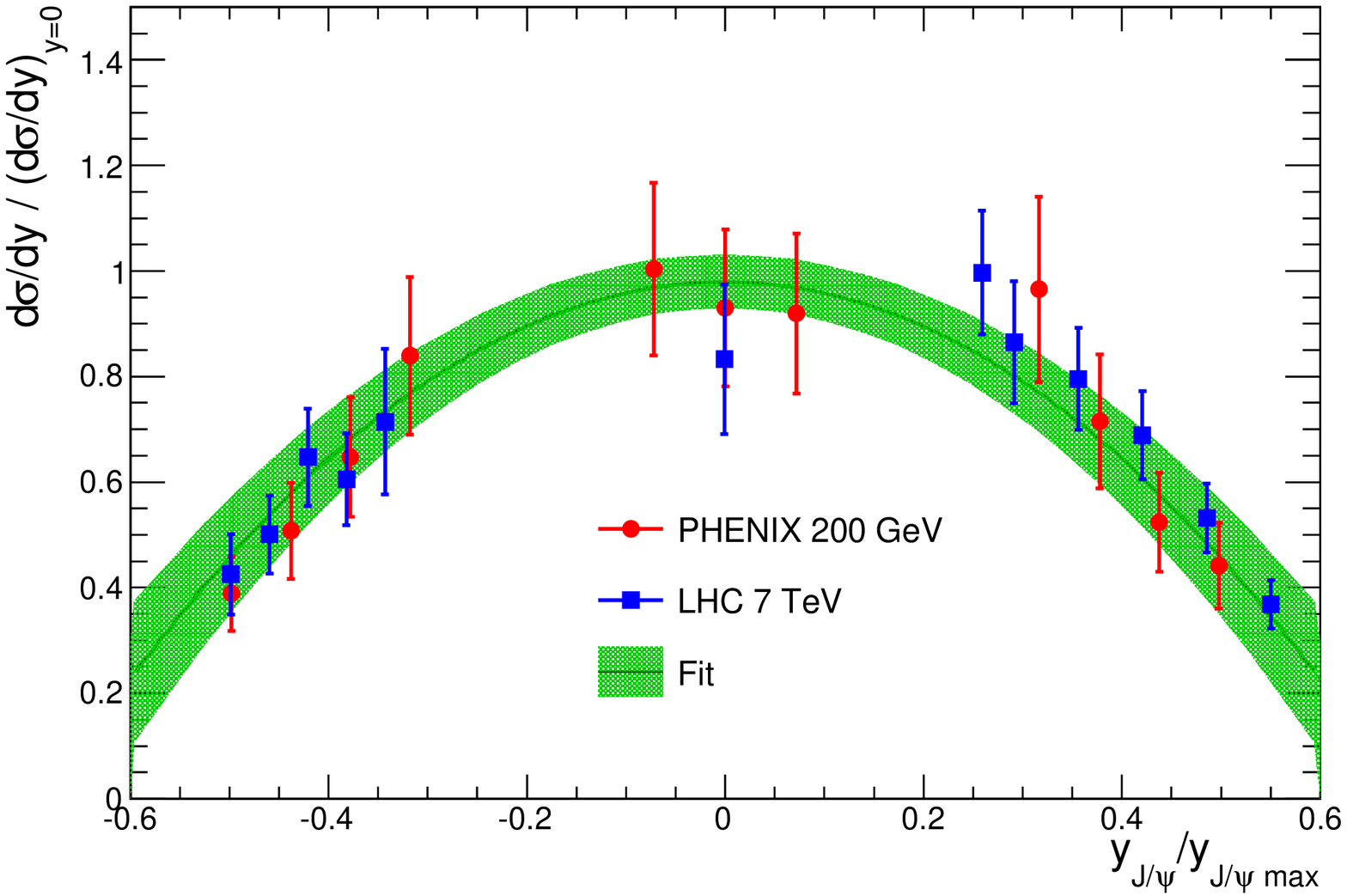}
           \label{fig:UniversalYreljpsi_Polya}
           }
  \subfigure[$n=4$]{
           \includegraphics[width=0.47\textwidth]{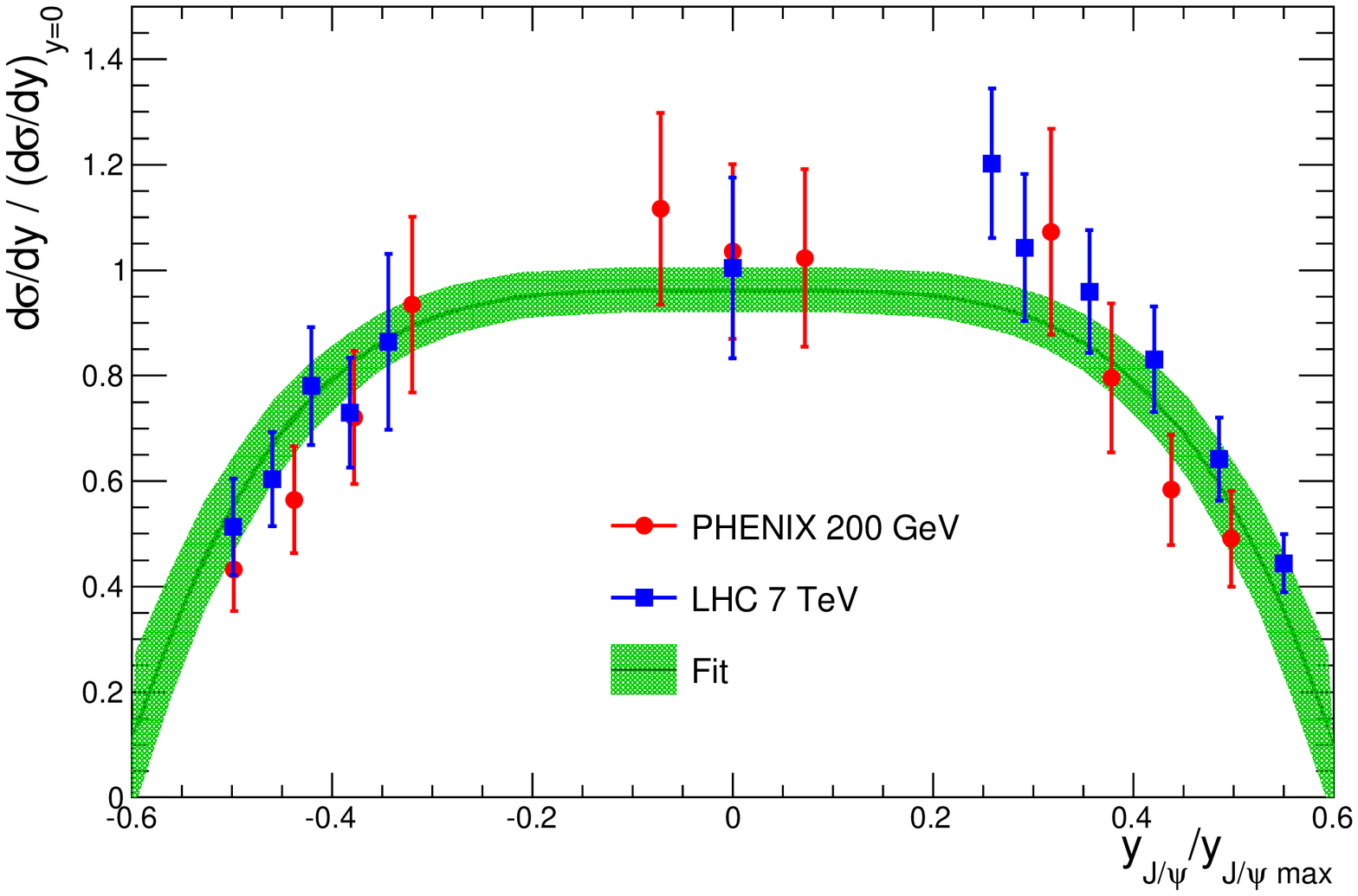}
           \label{fig:UniversalYreljpsi_Polyb}
           }
  \subfigure[$n=6$]{
           \includegraphics[width=0.47\textwidth]{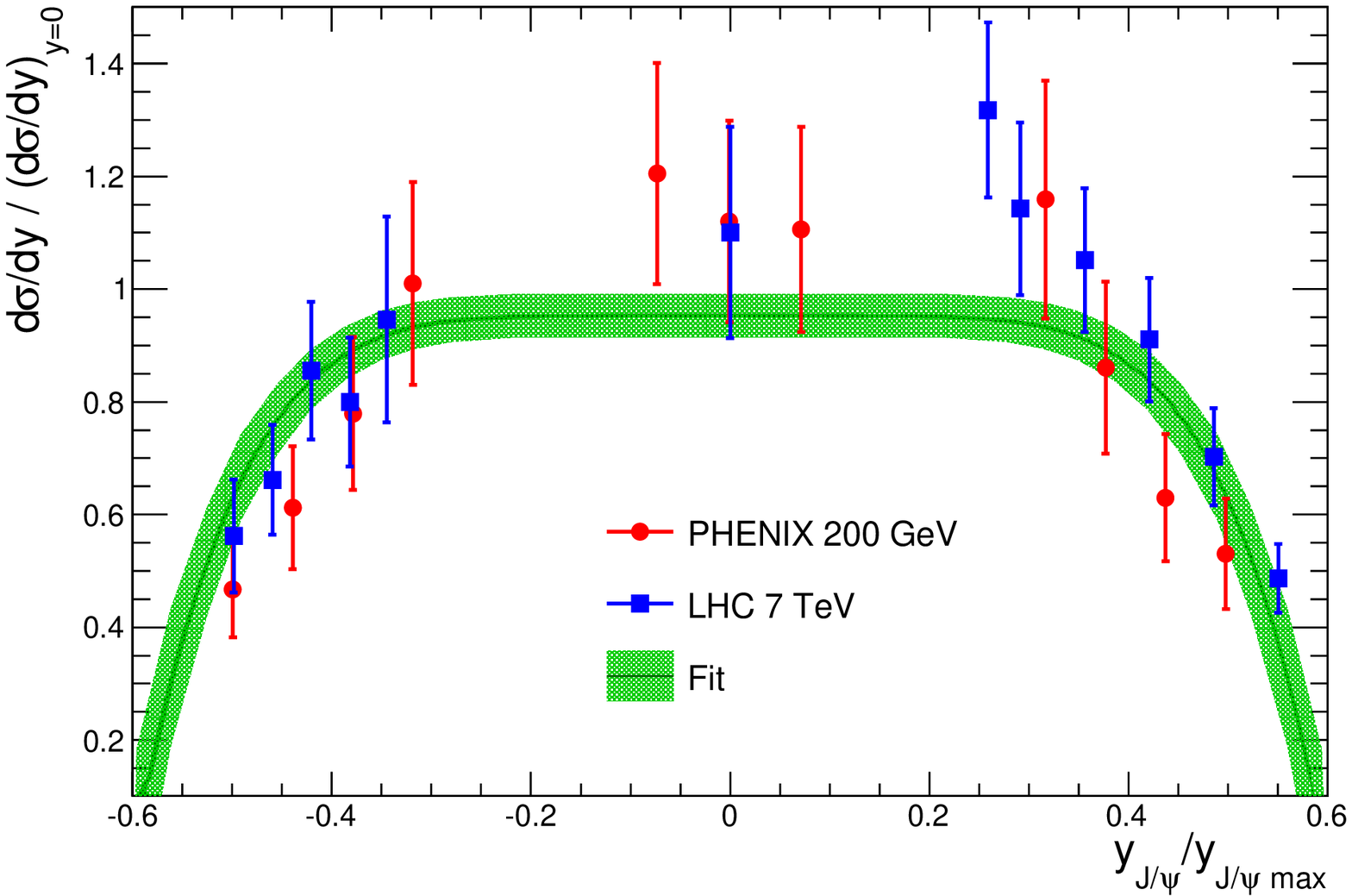}
           \label{fig:UniversalYreljpsi_Polyc}
           }
  \subfigure[$n=8$]{
           \includegraphics[width=0.47\textwidth]{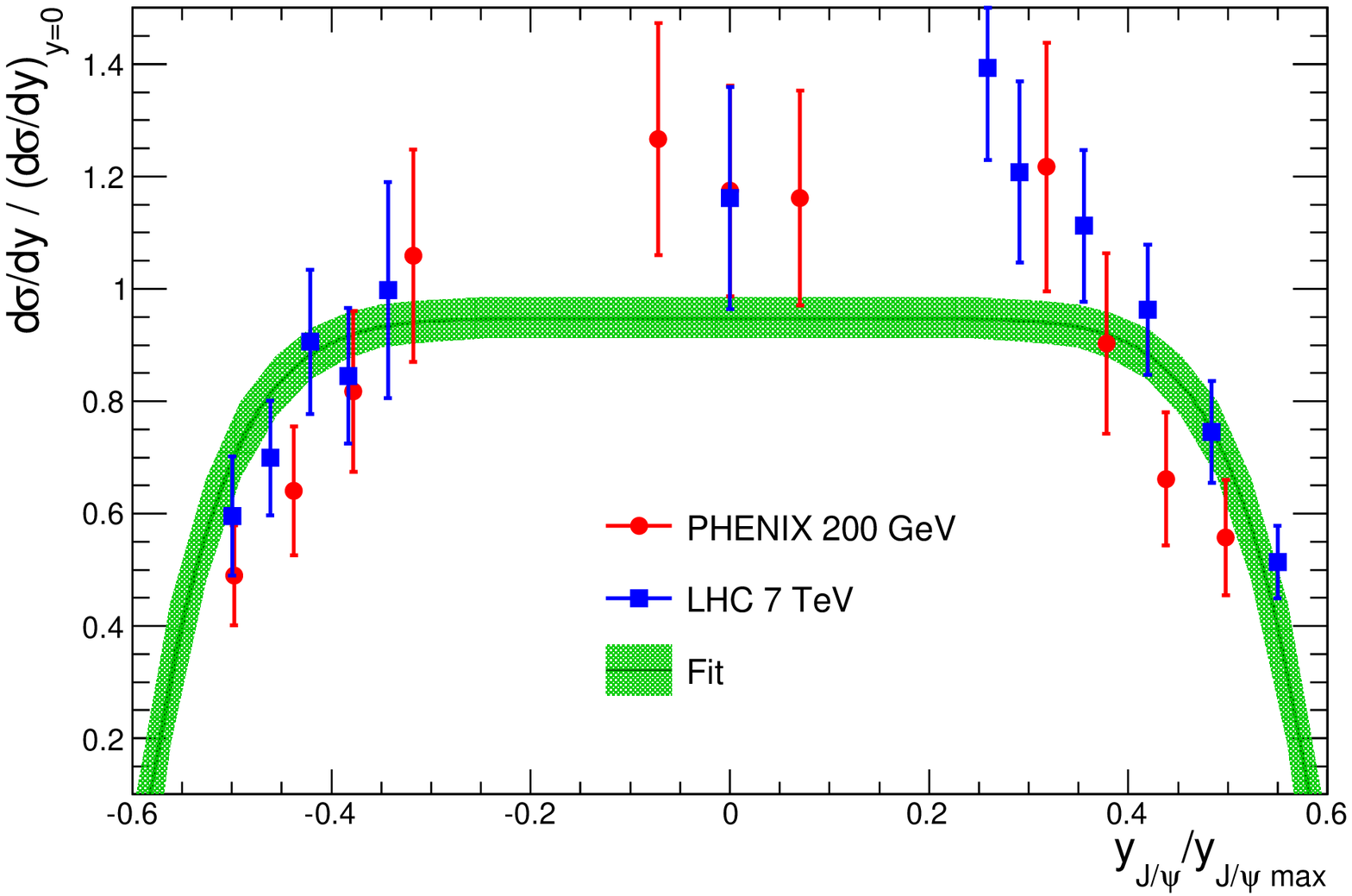}
           \label{fig:UniversalYreljpsi_Polyd}
           }
\caption{\label{fig:UniversalYreljpsi_Poly} 
	Inclusive \Jpsi production cross section as a function of the \Jpsi relative rapidity~($y /y_{\rm max}$) as fitted by a polynomial function, $ A - B \times (y / y_{\rm max})^n$ with $n=2$~\subref{fig:UniversalYreljpsi_Polya}, $4$~\subref{fig:UniversalYreljpsi_Polyb}, $6$~\subref{fig:UniversalYreljpsi_Polyc} and $8$~\subref{fig:UniversalYreljpsi_Polyd} 
in proton-proton collisions at 200 GeV (PHENIX~\cite{PHENIX07}) and at 7 TeV (ALICE, CMS LHCb~\cite{ALICEjpsi10, CMSjpsi10,LHCbjpsi10}).
        }
\end{figure}

We  compare the Gaussian fit results to those from a set of polynomial functions, defined as: 
\begin{equation}
 \frac{{\rm d}\sigma}{{\rm d}y} \, \,/  \,\, \frac{{\rm d}\sigma}{{\rm d}y} \Bigr|_{y=0}  =  A - B \times (y / y_{\rm max})^n 
\end{equation}
where $n$ is an even number (up to $8$). The resulting fits are shown in Fig.~\ref{fig:UniversalYreljpsi_Poly}. 
We deduce from the $\chi^2/$ndf values, reported in Table~\ref{tab:f_forw_pheno}, that all the functions are compatible with the measured $y$-differential distributions. The measurement uncertainties are not small enough to discriminate between quite different rapidity shapes, data driven extrapolations are affected by large uncertainties.

The scaling factor $f_{\rm forw}(y)$, defined in Eq.~\ref{eq:FforwardDef}, has been evaluated for all the considered functional form functions. The results are summarized in Table~\ref{tab:f_forw_pheno}. 
\begin{table}[!hbtp]
\begin{center}
\begin{tabular}{||l|cccc||} \hline \hline
Fit     & & $\chi^2$/ndf    & A     & B     \\   \hline
 Gaussian    & & 11.6/21 &\multicolumn{2}{c||}{($\sigma_y)~0.39\pm0.02$}   \\  
 Poly--(n=2) & & 9.2/21  &  $0.98 \pm 0.05$ & $2.1 \pm 0.2 $  \\
 Poly--(n=4) & & 17.3/21 &  $0.96 \pm 0.04$ & $6.6 \pm 0.8 $ \\
 Poly--(n=6) & & 27.5/21 &  $0.95 \pm 0.04$ & $20 \pm 3 $  \\
 Poly--(n=8) & & 36.3/21 &  $0.95 \pm 0.04$ & $65 \pm 9 $ \\
     \hline \hline
           & \multicolumn{4}{c||}{$f_{forw} (y)$} \\
Fit       & \multicolumn{2}{c}{$\sqrts=2.76$~TeV}  & \multicolumn{2}{c||}{$\sqrts=5.5$~TeV} \\  
            &  $y=2$  & $y=3.25$ & $y=2$ & $y=3.25$ \\  \hline
 Gaussian    
                       & $0.76 \pm 0.07$  & $0.48 \pm 0.07$  & $0.79 \pm 0.07$  & $0.54 \pm 0.07$  \\  
 Poly--(n=2) 
                       & $0.82 \pm 0.07$  & $0.51 \pm 0.11$  & $0.85 \pm 0.07$  & $0.60 \pm 0.10$  \\
 Poly--(n=4) 
                       & $0.95 \pm 0.05$  & $0.64 \pm 0.09$  & $0.97 \pm 0.05$  & $0.76 \pm 0.07$  \\
 Poly--(n=6) 
                       & $0.99 \pm 0.04$  & $0.78 \pm 0.07$  & $0.99 \pm 0.04$  & $0.74 \pm 0.06$  \\
 Poly--(n=8) 
                      & $1.00 \pm 0.04$  & $0.81 \pm 0.07$  & $1.00 \pm 0.04$  & $0.91 \pm 0.05$  \\[0.5ex] 
  Average     
  	&   $0.95^{+0.05}_{-0.19}$ &  $0.66^{+0.15}_{-0.18}$ &   $0.96^{+0.04}_{-0.17}$ &   $0.75^{+0.16}_{-0.21}$ \\[0.5ex]         
     \hline \hline
          & \multicolumn{4}{c||}{$BR_{ll} \times \left. \dsigmady \right|^{\rm incl.}_{y=y_0}$ [nb]} \\[0.5ex]
     $y_0$ & \multicolumn{2}{c}{$\sqrts=2.76$~TeV}  & \multicolumn{2}{c||}{$\sqrts=5.5$~TeV} \\[0.5ex]  \hline
     2.0 &  \multicolumn{2}{c}{$226 \pm 31~\textrm{\scriptsize (corr.)} \, ^{+13}_{-44}~\textrm{\scriptsize (uncorr.)}$} 
     	&  \multicolumn{2}{c||}{$334 \pm 69~\textrm{\scriptsize (corr.)} \, ^{+16}_{-58}~\textrm{\scriptsize (uncorr.)}$} \\[0.5ex]
     3.25 &  \multicolumn{2}{c}{$159 \pm 22~\textrm{\scriptsize (corr.)} \, ^{+35}_{-44}~\textrm{\scriptsize (uncorr.)}$} 
     	&  \multicolumn{2}{c||}{$264 \pm 54~\textrm{\scriptsize (corr.)} \, ^{+54}_{-75}~\textrm{\scriptsize (uncorr.)}$} \\[0.5ex] \hline \hline
\end{tabular}
\end{center}
\caption{Ratios of the mid- to forward-rapidity \Jpsi inclusive cross section scaling as obtained with different functional forms.
	The values of the fits $\chi^2$, the number of degrees of freedom (ndf), and the $A$ and $B$ parameters at $\sqrts=$200~GeV and 7~TeV are reported. 
	The interpolated y-differential production cross sections at $\sqrts=2.76$ and 5.5~TeV are also quoted.
	}
\label{tab:f_forw_pheno}
\label{tab:extrapToForwPheno}
\end{table}
Though it is observed that polynomial functions with $n=6$ and $n=8$ exhibit a larger $\chi^2$ than the ones with $n=2$ or $n=4$ and the Gaussian forms, the experimental measurements do not allow to disregard any fit function. We combine these results by computing their weighted average. The uncertainty is given by the envelope of their uncertainties. The results are reported in Table~\ref{tab:extrapToForwPheno}.
For completeness, a similar exercise was performed using the beam rapidity scaling ($y /y_{\rm beam \,\, max}$) instead of the \Jpsi relative rapidity ($y /y_{\rm max}$). The obtained forward rapidity scaling factors differ by less than 2\%. Since the difference is negligible with respect to the other uncertainties, we will neglect it. 
Using the factors $f_{\rm forw}$ reported here and our estimates of the mid-rapidity cross sections (Table~\ref{tab:final_midrap}), we obtain the inclusive differential cross sections quoted in Table~\ref{tab:extrapToForwPheno}. 
More precise measurements of the \Jpsi rapidity distribution would be needed to disentangle between different fit functions and reduce the interpolation uncertainties.  

\subsection{Discussion of the rapidity dependence results}
We observe that the results obtained with the functional form fits (Sec.~\ref{subsec:urp}) are in agreement with those obtained by the interpolation based on the FONLL and LO CEM calculations (Sec.~\ref{subsec:modelsInterpolation}). In all cases, the functional form approach involves larger uncertainties,  explained by the fact that it assumes no {\it a priori} knowledge of the rapidity shape. 
Here, we choose not to combine the results obtained with these two approaches.

%

\section{Transverse momentum distributions} 
\label{sec:ptinterpolation}

\begin{figure}[!hbtp]
   \centering
   \includegraphics[width=0.8\textwidth]{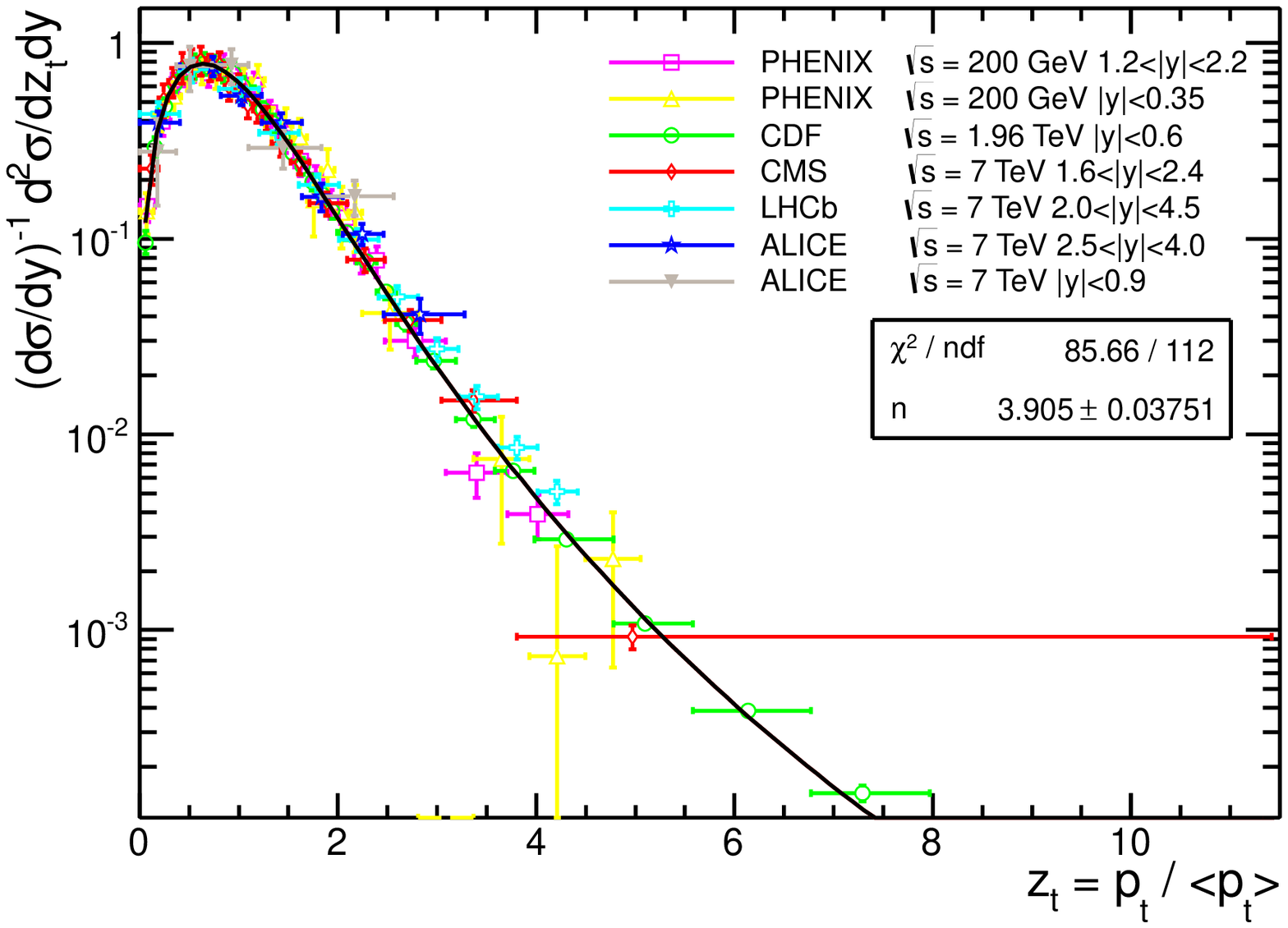}
   \caption{Universal $z_{\rm t}$ distribution for all the available experimental data.}
   \label{fig:UniversalpT}
\end{figure}

\begin{figure}[!hbtp]
   \centering
   \includegraphics[width=0.75\textwidth]{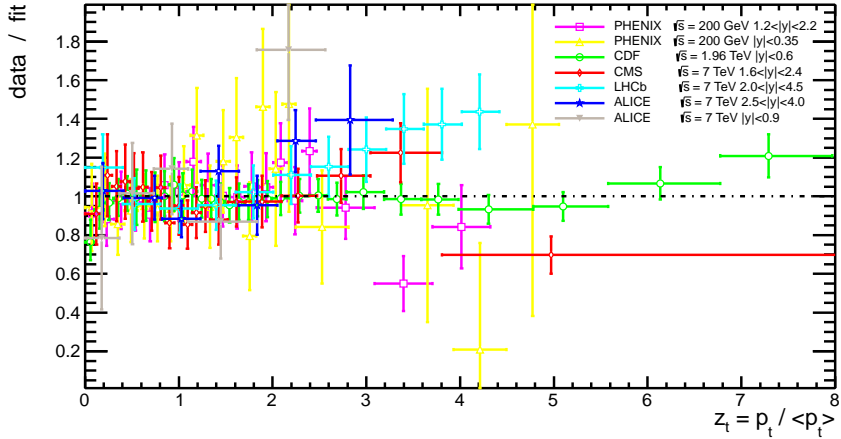}
   \caption{Ratio of the experimental data over the fit function shown in Fig.~\ref{fig:UniversalpT}.}
   \label{fig:DataOverFit}
\end{figure}

In this section we study the inclusive \Jpsi transverse momentum distribution. To compare the existing measurements at different energies and rapidity domains, we normalize their $\pt$-integrated cross section to unity and plot them together versus the $z_{\rm t}$ variable, defined as:
\begin{equation}
z_{\rm t}=\pt/\langle \pt \rangle,
\end{equation}

\noindent
where the values of $\langle \pt \rangle$ are given in Table~\ref{table1}. The resulting distributions (see Fig.~\ref{fig:UniversalpT}) show a universal behaviour which can be fitted by the following function:
\begin{equation}
\label{eq:scalingEq}
\frac{1}{{\rm d}\sigma/{\rm d}y} \, \frac{{\rm d}^2\sigma}{{\rm d}z_{\rm t} \, {\rm d}y} = c \times \frac{z_{\rm t}}{(1+a^2 z_{\rm t}^2)^n}, 
\end{equation}
where $c = 2 \, a^2 \, (n-1)$ follows from the normalization to unity. Requiring $\langle z_{\rm t} \rangle = 1$, meaning that $\langle \pt \rangle$ computed by the fit function is equal to that of Table~\ref{table1}, results in: 
$a = \Gamma(3/2) \, \Gamma(n-3/2) \, / \, \Gamma(n-1)$. 
We are then left with only one free parameter $n$. We do a global fit to all the experimental data and obtain: $n=3.9$ (see Fig.~\ref{fig:UniversalpT}). The systematic uncertainties on the $n$ parameter are evaluated by performing three global-like fits excluding one by one the data at 200~GeV, 1.96~TeV and 7~TeV. This gives finally: $n=3.9^{+0.5}_{-0.1}$. To demonstrate the scaling versus $z_{\rm t}$ variable in more details, we plot in Fig.~\ref{fig:DataOverFit} the ratio of the experimental data to the fit function value at each data point. We have checked that this fit function describes well also the prompt \Jpsi data at 7 TeV~\cite{CMSjpsi10,LHCbjpsi10}. Similar fit of the $z_{\rm t}$ scaling function Eq.~(\ref{eq:scalingEq}) to the $\Upsilon (1S)$ meson $\pt$ spectra measured at 1.8 TeV~\cite{CDFupsi02} and 7 TeV~\cite{CMSupsi10,LHCbupsi10} gives $n=3.4$.

\begin{table}[!hbtp]
\begin{center}
\begin{tabular}{||c|lccccc||} \hline \hline
& Data excluded  & \multicolumn{5}{c||}{$\sqrts$~[TeV]} \\ 
 & & 0.2 & 1.96  & 2.76 & 5.5 & 7 \\  \hline
 & none &  &  & 2.31 & 2.48 &  \\
$\langle \pt \rangle$ & RHIC & 2.52 &  & 2.51 & 2.50 &  \\
$[$GeV/$c]$  & Tevatron &  & 2.17 & 2.25 & 2.44 &  \\
& LHC &  &  & 2.67 &3.01 & 3.15\\ \hline
Syst. Unc. & & 48\% & 13\% & 12\% & 17\%  & 25\% \\ \hline \hline
\end{tabular}
\end{center}
\caption{Inclusive \Jpsi $\langle \pt \rangle$ versus $\sqrts$ and its systematic uncertainty (Syst. Unc.).}
\label{tab: meanpT}
\end{table}

The next step to obtain the $\pt$ distributions at 2.76 and 5.5 TeV is to estimate the \Jpsi $\langle \pt \rangle$ at these two energies. For this we fit the PHENIX, CDF, CMS, ALICE and LHCb $\langle \pt \rangle$ data by a power-law function of energy. The results are reported in Table~\ref{tab: meanpT}. To evaluate the calculation systematics, data points at different energies have been removed one by one. The last row of Table~\ref{tab: meanpT} presents the systematic uncertainty on $\langle \pt \rangle$, estimated as the difference between the extrapolated and measured values for $\sqrts$ = 200 GeV, 1.96 TeV and 7 TeV, and the spread of the results from the global fit and the fits excluding data points for $\sqrts$ = 2.76~TeV and 5.5~TeV.

\begin{figure}[!hbtp]
   \centering
  \subfigure[$\sqrts=2.76$~TeV]{
   \includegraphics[width=0.62\textwidth]{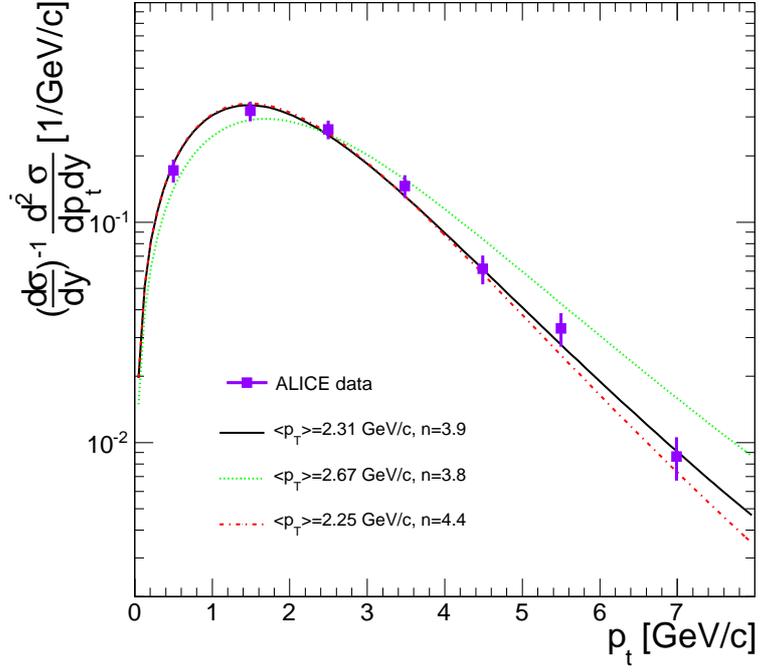}
      \label{fig:pT2760}
   }
  \subfigure[$\sqrts=5.5$~TeV]{
   \includegraphics[width=0.62\textwidth]{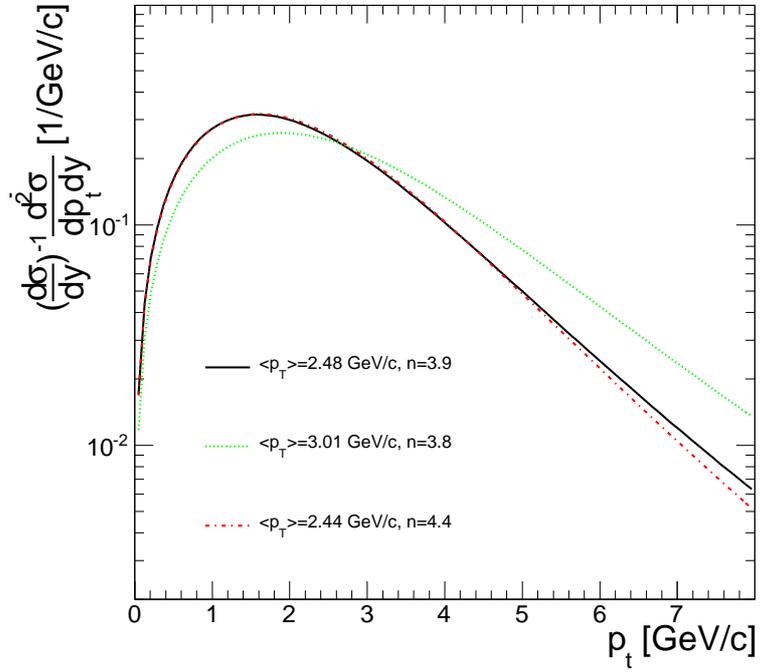}
   \label{fig:pT5500}
   }
   \caption{Extrapolated normalized $\pt$ distribution at 2.76~TeV~\subref{fig:pT2760} and 5.5~TeV~\subref{fig:pT5500} with the uncertainty bands as compared with the ALICE data at $\sqrts=2.76$~TeV.}
   \label{fig:pTextap}
\end{figure}

The $\pt$ distribution interpolation at 2.76 and 5.5 TeV is done using Eq.~(\ref{eq:scalingEq}) and the corresponding $\langle \pt \rangle$. The resulting central distributions and their uncertainty bands are shown in Fig.~\ref{fig:pT2760} and Fig.~\ref{fig:pT5500} respectively.
%

%

\section{Comparison with the ALICE measurements at $\sqrts=2.76$~TeV}
\label{sec:compWithAlice}
The ALICE collaboration did recently measure the inclusive \Jpsi production at $\sqrts=2.76$~TeV~\cite{Arnaldi11}. 
The inclusive \Jpsi production cross section at mid-rapidity, $BR_{ll} \times \dsigmady|_{y=0} = 221 \pm 57$~nb~\cite{Arnaldi11}, is in good agreement with the interpolation value, $239^{+6}_{-10}~\textrm{(model)} \pm 31~\textrm{(fit)}$~nb, see Table~\ref{tab:final_midrap}. 
The $\pt$-integrated rapidity distribution, see Fig.~\ref{fig:rapidityJpsi276}, is well described by our interpolation. 
The cross section measurements suggest that the mid- to forward-rapidity factor is: $f_{forw}^{\rm ALICE}(-3.25<y<-3.0)= 0.6 \pm 0.2$, and $f_{forw}^{\rm ALICE}(-3.5<y<-3.25)= 0.5 \pm 0.2$. 
Comparing these values to the interpolation results reported in Tables~\ref{tab:test_rap}--\ref{tab:extrapToForwPheno}, we observe that the theory driven (FONLL and LO CEM) and functional form (Gaussian, polynomial-$n=2$, polynomial-$n=4$) calculations reproduce the measured trend. However, the polynomial-$n=6$ and polynomial-$n=8$ functional forms tend to overestimate the measured cross sections at forward-rapidities. The ALICE results at $\sqrts=2.76$~TeV~\cite{Arnaldi11} seem then to advocate in favor of neglecting the polynomial-$n=6$ and polynomial-$n=8$ functions of the rapidity interpolation approach for future calculations. 
If we disregard these two functions, the functional form rapidity interpolation factors at $\sqrts=2.76$~TeV would be of 
$f_{forw}(y=3.25) = 0.53^{+0.11}_{-0.05}$ and  
$f_{forw}(y=2.0) = 0.87^{+0.08}_{-0.11}$, 
while at $\sqrts=5.5$~TeV they would be 
$f_{forw}(y=3.25) = 0.64^{+0.12}_{-0.10}$ and  
$f_{forw}(y=2.0) = 0.89^{+0.07}_{-0.10}$, to replace those of Table~\ref{tab:extrapToForwPheno}.
The measured transverse momentum distribution, Fig.~\ref{fig:pT2760}, is well reproduced by our predictions.

%

\section{Conclusions}
\label{sec:conclusions}

A procedure to evaluate the energy dependence of the inclusive (and possibly prompt) \Jpsi integrated and differential cross sections in proton proton collisions has been established and used to compute the \Jpsi cross sections at the same centre of mass energy as the one available (per nucleon pair) in \PbPb~collisions at the LHC (2.76~TeV at present and 5.5~TeV in the near future). 
The calculation exploits the available experimental measurements in three distinct steps. 
The first step consists in describing the energy dependence of the mid-rapidity $\pt$-integrated cross section~(Sec.~\ref{sec:energyinterpolation}). The expectations from functional form (power-law) and model-driven (FONLL and LO CEM) fits have been combined together. The results are summarized in Table~\ref{tab:summary}. The interpolation systematics is of 15\%~(20\%) at 2.76~(5.5)~TeV. 
The second step is the determination of the scaling factors from mid- to forward rapidities (Sec.~\ref{sec:rapidityinterpolation}). 
Two approaches have been discussed, one based on the theoretically (pQCD) predicted rapidity shape, the other being a functional form approach, based on the energy scaling of the $y /y_{\rm max}$ distributions. The results with both methods are compatible, although the functional form approach involves larger uncertainties. Higher precision measurements are needed in order to better constrain the rapidity distribution. 
The third step is the description of the energy evolution of the transverse momentum distribution (Sec.~\ref{sec:ptinterpolation}), obtained by assuming an universal scaling of the $z_{\rm t}$ distributions (Eq.~\ref{eq:scalingEq}). The predicted distributions at $\sqrts=2.76$ and 5.5~TeV are drawn in Figs.~\ref{fig:pT2760} and~\ref{fig:pT5500}. 
Our predictions of the $\pt$-integrated and differential cross sections at $\sqrts=2.76$~TeV are in agreement, within interpolation and experimental uncertainties, with the ALICE measurements at the same energy. 
These comparisons confirm the appropriateness of our interpolation procedure, that can be used to obtain a \pp reference for the analysis of both A--A and p--A data at the LHC.
\begin{table}[!htbp]
\begin{center}
\begin{tabular}{||c|cc||} \hline \hline
 $y_0$ & \multicolumn{2}{c||}{$BR_{ll} \times \left. \dsigmady \right|^{\rm incl.}_{y=y_0}$~[nb]}  \\[0.5ex]  
        &      $\sqrts=2.76$x~TeV                                   &                   $\sqrts=5.5$~TeV                  \\[0.5ex]  \hline
   0    & $239 ^{+6}_{-10}~\textrm{\scriptsize (model)} \pm 31~\textrm{\scriptsize (fit)}$    
   	& $350 ^{+20}_{-51}~\textrm{\scriptsize (model)} \pm 51~\textrm{\scriptsize (fit)}$ \\[0.5ex]
  2.0   & $208 \pm 28~\textrm{\scriptsize (corr.)} ^{+9}_{-6}~\textrm{\scriptsize (uncorr.)}$   
  	& $316 \pm 65~\textrm{\scriptsize (corr.)} ^{+14}_{-7}~\textrm{\scriptsize (uncorr.)}$      \\[0.5ex]
  3.25  & $153 \pm 21~\textrm{\scriptsize (corr.)} ^{+12}_{-12}~\textrm{\scriptsize (uncorr.)}$ 
  	& $256 \pm 53~\textrm{\scriptsize (corr.)} ^{+23}_{-16}~\textrm{\scriptsize (uncorr.)}$      \\[0.5ex]
    \hline \hline
     \end{tabular}
\end{center}
\caption{Summary of the interpolation results for the inclusive \Jpsi production cross sections at specific rapidity values for $\sqrts=2.76$ and 5.5~TeV. }
\label{tab:summary}
\end{table}

\section*{Acknowledgements} 

The authors would like to express their gratitude to J.~Schukraft for careful reading and valuable suggestions. Our acknowledgement also goes to D.~Stocco, for his help on the LO CEM implementation. 
Part of this work was supported by the ReteQuarkonii Networking of EU I3 Hadron Physics Program. 
The work of S.G. was supported partially by the CNRS-RFBR PICS project number 4870.


\bibliographystyle{elsarticle-num}



\end{document}